\documentclass[a4paper,12pt]{article}
\usepackage{amsfonts}
\usepackage{amsmath}
\usepackage{amssymb}
\usepackage{graphicx}
\usepackage[english]{babel}

\makeatletter
 
\def\diagram{\m@th\leftwidth=\z@ \rightwidth=\z@ \topheight=\z@
\botheight=\z@ \setbox\@picbox\hbox\bgroup}
 
\def\enddiagram{\egroup\wd\@picbox\rightwidth\unitlength
\ht\@picbox\topheight\unitlength \dp\@picbox\botheight\unitlength
\hskip\leftwidth\unitlength\box\@picbox}
 
\def\bfig{\begin{diagram}}
\def\efig{\end{diagram}}
\newcount\wideness \newcount\leftwidth \newcount\rightwidth
\newcount\highness \newcount\topheight \newcount\botheight
 
\def\ratchet#1#2{\ifnum#1<#2 \global #1=#2 \fi}
 
\def\putbox(#1,#2)#3{%
\horsize{\wideness}{#3} \divide\wideness by 2
{\advance\wideness by #1 \ratchet{\rightwidth}{\wideness}}
{\advance\wideness by -#1 \ratchet{\leftwidth}{\wideness}}
\vertsize{\highness}{#3} \divide\highness by 2
{\advance\highness by #2 \ratchet{\topheight}{\highness}}
{\advance\highness by -#2 \ratchet{\botheight}{\highness}}
\put(#1,#2){\makebox(0,0){$#3$}}}
 
\def\putlbox(#1,#2)#3{%
\horsize{\wideness}{#3}
{\advance\wideness by #1 \ratchet{\rightwidth}{\wideness}}
{\ratchet{\leftwidth}{-#1}}
\vertsize{\highness}{#3} \divide\highness by 2
{\advance\highness by #2 \ratchet{\topheight}{\highness}}
{\advance\highness by -#2 \ratchet{\botheight}{\highness}}
\put(#1,#2){\makebox(0,0)[l]{$#3$}}}
 
\def\putrbox(#1,#2)#3{%
\horsize{\wideness}{#3}
{\ratchet{\rightwidth}{#1}}
{\advance\wideness by -#1 \ratchet{\leftwidth}{\wideness}}
\vertsize{\highness}{#3} \divide\highness by 2
{\advance\highness by #2 \ratchet{\topheight}{\highness}}
{\advance\highness by -#2 \ratchet{\botheight}{\highness}}
\put(#1,#2){\makebox(0,0)[r]{$#3$}}}

\def\adjust[#1]{} 
 
\newcount \coefa
\newcount \coefb
\newcount \coefc
\newcount\tempcounta
\newcount\tempcountb
\newcount\tempcountc
\newcount\tempcountd
\newcount\xext
\newcount\yext
\newcount\xoff
\newcount\yoff
\newcount\gap%
\newcount\arrowtypea
\newcount\arrowtypeb
\newcount\arrowtypec
\newcount\arrowtyped
\newcount\arrowtypee
\newcount\height
\newcount\width
\newcount\xpos
\newcount\ypos
\newcount\run
\newcount\rise
\newcount\arrowlength
\newcount\halflength
\newcount\arrowtype
\newdimen\tempdimen
\newdimen\xlen
\newdimen\ylen
\newsavebox{\tempboxa}%
\newsavebox{\tempboxb}%
\newsavebox{\tempboxc}%
 
\newdimen\w@dth
 
\def\setw@dth#1#2{\setbox\z@\hbox{\m@th$#1$}\w@dth=\wd\z@
\setbox\@ne\hbox{\m@th$#2$}\ifnum\w@dth<\wd\@ne \w@dth=\wd\@ne \fi
\advance\w@dth by 1.2em}
 
\def\t@^#1_#2{\allowbreak\def\n@one{#1}\def\n@two{#2}\mathrel
{\setw@dth{#1}{#2}
\mathop{\hbox to \w@dth{\rightarrowfill}}\limits
\ifx\n@one\empty\else ^{\box\z@}\fi
\ifx\n@two\empty\else _{\box\@ne}\fi}}
\def\t@@^#1{\@ifnextchar_{\t@^{#1}}{\t@^{#1}_{}}}
\def\to{\@ifnextchar^{\t@@}{\t@@^{}}}
 
\def\t@left^#1_#2{\def\n@one{#1}\def\n@two{#2}\mathrel{\setw@dth{#1}{#2}
\mathop{\hbox to \w@dth{\leftarrowfill}}\limits
\ifx\n@one\empty\else ^{\box\z@}\fi
\ifx\n@two\empty\else _{\box\@ne}\fi}}
\def\t@@left^#1{\@ifnextchar_{\t@left^{#1}}{\t@left^{#1}_{}}}
\def\toleft{\@ifnextchar^{\t@@left}{\t@@left^{}}}
 
\def\two@^#1_#2{\allowbreak
\def\n@one{#1}\def\n@two{#2}\mathrel{\setw@dth{#1}{#2}
\mathop{\vcenter{\lineskip\z@\baselineskip\z@
                 \hbox to \w@dth{\rightarrowfill}%
                 \hbox to \w@dth{\rightarrowfill}}%
       }\limits
\ifx\n@one\empty\else ^{\box\z@}\fi
\ifx\n@two\empty\else _{\box\@ne}\fi}}
\def\tw@@^#1{\@ifnextchar _{\two@^{#1}}{\two@^{#1}_{}}}
\def\two{\@ifnextchar ^{\tw@@}{\tw@@^{}}}
 
\def\tofr@^#1_#2{\def\n@one{#1}\def\n@two{#2}\mathrel{\setw@dth{#1}{#2}
\mathop{\vcenter{\hbox to \w@dth{\rightarrowfill}\kern-1.7ex
                 \hbox to \w@dth{\leftarrowfill}}%
       }\limits
\ifx\n@one\empty\else ^{\box\z@}\fi
\ifx\n@two\empty\else _{\box\@ne}\fi}}
\def\t@fr@^#1{\@ifnextchar_ {\tofr@^{#1}}{\tofr@^{#1}_{}}}
\def\tofro{\@ifnextchar^ {\t@fr@}{\t@fr@^{}}}

\def\mon{\mathop{\m@th\hbox to
      14.6\P@{\lasyb\char'51\hskip-2.1\P@$\arrext$\hss
$\mathord\rightarrow$}}\limits} 
\def\leftmono{\mathrel{\m@th\hbox to
14.6\P@{$\mathord\leftarrow$\hss$\arrext$\hskip-2.1\P@\lasyb\char'50%
}}\limits} 
\mathchardef\arrext="0200       

\def\settypes(#1,#2,#3){\arrowtypea#1 \arrowtypeb#2 \arrowtypec#3}
\def\settoheight#1#2{\setbox\@tempboxa\hbox{#2}#1\ht\@tempboxa\relax}%
\def\settodepth#1#2{\setbox\@tempboxa\hbox{#2}#1\dp\@tempboxa\relax}%
\def\settokens`#1`#2`#3`#4`{%
     \def\tokena{#1}\def\tokenb{#2}\def\tokenc{#3}\def\tokend{#4}}
\def\setsqparms[#1`#2`#3`#4;#5`#6]{%
\arrowtypea #1
\arrowtypeb #2
\arrowtypec #3
\arrowtyped #4
\width #5
\height #6
}
\def\setpos(#1,#2){\xpos=#1 \ypos#2}

\def\settriparms[#1`#2`#3;#4]{\settripairparms[#1`#2`#3`1`1;#4]}%
 
\def\settripairparms[#1`#2`#3`#4`#5;#6]{%
\arrowtypea #1
\arrowtypeb #2
\arrowtypec #3
\arrowtyped #4
\arrowtypee #5
\width #6
\height #6
}
 
\def\resetparms{\settripairparms[1`1`1`1`1;500]\width 500}
 
\resetparms
 
\def\mvector(#1,#2)#3{
\put(0,0){\vector(#1,#2){#3}}%
\put(0,0){\vector(#1,#2){26}}%
}
\def\evector(#1,#2)#3{{
\arrowlength #3
\put(0,0){\vector(#1,#2){\arrowlength}}%
\advance \arrowlength by-30
\put(0,0){\vector(#1,#2){\arrowlength}}%
}}
 
\def\horsize#1#2{%
\settowidth{\tempdimen}{$#2$}%
#1=\tempdimen
\divide #1 by\unitlength
}
 
\def\vertsize#1#2{%
\settoheight{\tempdimen}{$#2$}%
#1=\tempdimen
\settodepth{\tempdimen}{$#2$}%
\advance #1 by\tempdimen
\divide #1 by\unitlength
}
 
\def\putvector(#1,#2)(#3,#4)#5#6{{%
\ifnum3<\arrowtype
\putdashvector(#1,#2)(#3,#4)#5\arrowtype
\else
\ifnum\arrowtype<-3
\putdashvector(#1,#2)(#3,#4)#5\arrowtype
\else
\xpos=#1
\ypos=#2
\run=#3
\rise=#4
\arrowlength=#5
\ifnum \arrowtype<0
    \ifnum \run=0
        \advance \ypos by-\arrowlength
    \else
        \tempcounta \arrowlength
        \multiply \tempcounta by\rise
        \divide \tempcounta by\run
        \ifnum\run>0
            \advance \xpos by\arrowlength
            \advance \ypos by\tempcounta
        \else
            \advance \xpos by-\arrowlength
            \advance \ypos by-\tempcounta
        \fi
    \fi
    \multiply \arrowtype by-1
    \multiply \rise by-1
    \multiply \run by-1
\fi
\ifcase \arrowtype
\or \put(\xpos,\ypos){\vector(\run,\rise){\arrowlength}}%
\or \put(\xpos,\ypos){\mvector(\run,\rise)\arrowlength}%
\or \put(\xpos,\ypos){\evector(\run,\rise){\arrowlength}}%
\fi\fi\fi
}}
 
\def\putsplitvector(#1,#2)#3#4{
\xpos #1
\ypos #2
\arrowtype #4
\halflength #3
\arrowlength #3
\gap 140
\advance \halflength by-\gap
\divide \halflength by2
\ifnum\arrowtype>0
   \ifcase \arrowtype
   \or \put(\xpos,\ypos){\line(0,-1){\halflength}}%
       \advance\ypos by-\halflength
       \advance\ypos by-\gap
       \put(\xpos,\ypos){\vector(0,-1){\halflength}}%
   \or \put(\xpos,\ypos){\line(0,-1)\halflength}%
       \put(\xpos,\ypos){\vector(0,-1)3}%
       \advance\ypos by-\halflength
       \advance\ypos by-\gap
       \put(\xpos,\ypos){\vector(0,-1){\halflength}}%
   \or \put(\xpos,\ypos){\line(0,-1)\halflength}%
       \advance\ypos by-\halflength
       \advance\ypos by-\gap
       \put(\xpos,\ypos){\evector(0,-1){\halflength}}%
   \fi
\else \arrowtype=-\arrowtype
   \ifcase\arrowtype
   \or \advance \ypos by-\arrowlength
       \put(\xpos,\ypos){\line(0,1){\halflength}}%
       \advance\ypos by\halflength
       \advance\ypos by\gap
       \put(\xpos,\ypos){\vector(0,1){\halflength}}%
   \or \advance \ypos by-\arrowlength
       \put(\xpos,\ypos){\line(0,1)\halflength}%
       \put(\xpos,\ypos){\vector(0,1)3}%
       \advance\ypos by\halflength
       \advance\ypos by\gap
       \put(\xpos,\ypos){\vector(0,1){\halflength}}%
   \or \advance \ypos by-\arrowlength
       \put(\xpos,\ypos){\line(0,1)\halflength}%
       \advance\ypos by\halflength
       \advance\ypos by\gap
       \put(\xpos,\ypos){\evector(0,1){\halflength}}%
   \fi
\fi
}
 
\def\putmorphism(#1)(#2,#3)[#4`#5`#6]#7#8#9{{%
\run #2
\rise #3
\ifnum\rise=0
  \puthmorphism(#1)[#4`#5`#6]{#7}{#8}#9%
\else\ifnum\run=0
  \putvmorphism(#1)[#4`#5`#6]{#7}{#8}#9%
\else
\setpos(#1)%
\arrowlength #7
\arrowtype #8
\ifnum\run=0
\else\ifnum\rise=0
\else
\ifnum\run>0
    \coefa=1
\else
   \coefa=-1
\fi
\ifnum\arrowtype>0
   \coefb=0
   \coefc=-1
\else
   \coefb=\coefa
   \coefc=1
   \arrowtype=-\arrowtype
\fi
\width=2
\multiply \width by\run
\divide \width by\rise
\ifnum \width<0  \width=-\width\fi
\advance\width by60
\if l#9 \width=-\width\fi
\putbox(\xpos,\ypos){#4}
{\multiply \coefa by\arrowlength
\advance\xpos by\coefa
\multiply \coefa by\rise
\divide \coefa by\run
\advance \ypos by\coefa
\putbox(\xpos,\ypos){#5} }%
{\multiply \coefa by\arrowlength
\divide \coefa by2
\advance \xpos by\coefa
\advance \xpos by\width
\multiply \coefa by\rise
\divide \coefa by\run
\advance \ypos by\coefa
\if l#9%
   \putrbox(\xpos,\ypos){#6}%
\else\if r#9%
   \putlbox(\xpos,\ypos){#6}%
\fi\fi }%
{\multiply \rise by-\coefc
\multiply \run by-\coefc
\multiply \coefb by\arrowlength
\advance \xpos by\coefb
\multiply \coefb by\rise
\divide \coefb by\run
\advance \ypos by\coefb
\multiply \coefc by70
\advance \ypos by\coefc
\multiply \coefc by\run
\divide \coefc by\rise
\advance \xpos by\coefc
\multiply \coefa by140
\multiply \coefa by\run
\divide \coefa by\rise
\advance \arrowlength by\coefa
\ifcase\arrowtype
\or \put(\xpos,\ypos){\vector(\run,\rise){\arrowlength}}%
\or \put(\xpos,\ypos){\mvector(\run,\rise){\arrowlength}}%
\or \put(\xpos,\ypos){\evector(\run,\rise){\arrowlength}}%
\fi}\fi\fi\fi\fi}}

\newcount\numbdashes \newcount\lengthdash \newcount\increment
 
\def\howmanydashes{
\numbdashes=\arrowlength \lengthdash=40
\divide\numbdashes by \lengthdash
\lengthdash=\arrowlength
\divide\lengthdash by \numbdashes
\increment=\lengthdash
\multiply\lengthdash by 3
\divide\lengthdash by 5
}
 
\def\putdashvector(#1)(#2,#3)#4#5{%
\ifnum#3=0 \putdashhvector(#1){#4}#5
\else
\ifnum#2=0
\putdashvvector(#1){#4}#5\fi\fi}
 
\def\putdashhvector(#1,#2)#3#4{{%
\arrowlength=#3 \howmanydashes
\multiput(#1,#2)(\increment,0){\numbdashes}%
{\vrule height .4pt width \lengthdash\unitlength}
\arrowtype=#4 \xpos=#1
\ifnum\arrowtype<0 \advance\arrowtype by 7 \fi
\ifcase\arrowtype
\or \advance\xpos by 10
    \put(\xpos,#2){\vector(-1,0){\lengthdash}}
    \advance\xpos by 40
    \put(\xpos,#2){\vector(-1,0){\lengthdash}}
\or \advance \xpos by 10
    \put(\xpos,#2){\vector(-1,0){\lengthdash}}
    \advance\xpos by  \arrowlength
    \advance\xpos by  -50
    \put(\xpos,#2){\vector(-1,0){\lengthdash}}
\or \advance\xpos by 10
    \put(\xpos,#2){\vector(-1,0){\lengthdash}}
\or \advance\xpos by \arrowlength
    \advance\xpos by -\lengthdash
    \put(\xpos,#2){\vector(1,0){\lengthdash}}
\or {\advance\xpos by 10
    \put(\xpos,#2){\vector(1,0){\lengthdash}}}
    \advance\xpos by \arrowlength
    \advance\xpos by -\lengthdash
    \put(\xpos,#2){\vector(1,0){\lengthdash}}
\or \advance\xpos by \arrowlength
    \advance\xpos by -\lengthdash
    \put(\xpos,#2){\vector(1,0){\lengthdash}}
    \advance\xpos by -40
    \put(\xpos,#2){\vector(1,0){\lengthdash}}
   \fi
}}
 
\def\putdashvvector(#1,#2)#3#4{{%
\arrowlength=#3 \howmanydashes
\ypos=#2 \advance\ypos by -\arrowlength
\multiput(#1,#2)(0,\increment){\numbdashes}%
    {\vrule width .4pt height \lengthdash\unitlength}
\arrowtype=#4 \ypos=#2
\ifnum\arrowtype<0 \advance\arrowtype by 7 \fi
\ifcase\arrowtype
\or \advance\ypos by \arrowlength \advance\ypos by -40
    \put(#1,\ypos){\vector(0,1){\lengthdash}}
    \advance\ypos by -40
    \put(#1,\ypos){\vector(0,1){\lengthdash}}
\or \advance\ypos by 10
    \put(#1,\ypos){\vector(0,1){\lengthdash}}
    \advance\ypos by \arrowlength \advance\ypos by -40
    \put(#1,\ypos){\vector(0,1){\lengthdash}}
\or \advance\ypos by \arrowlength \advance\ypos by -40
    \put(#1,\ypos){\vector(0,1){\lengthdash}}
\or \advance\ypos by 10
    \put(#1,\ypos){\vector(0,-1){\lengthdash}}
\or \advance\ypos by 10
    \put(#1,\ypos){\vector(0,-1){\lengthdash}}
    \advance\ypos by \arrowlength \advance\ypos by -40
    \put(#1,\ypos){\vector(0,-1){\lengthdash}}
\or \advance\ypos by 10
    \put(#1,\ypos){\vector(0,-1){\lengthdash}}
    \advance\ypos by 40
    \put(#1,\ypos){\vector(0,-1){\lengthdash}}
\fi
}}
 
\def\puthmorphism(#1,#2)[#3`#4`#5]#6#7#8{{%
\xpos #1
\ypos #2
\width #6
\arrowlength #6
\arrowtype=#7
\putbox(\xpos,\ypos){#3\vphantom{#4}}%
{\advance \xpos by\arrowlength
\putbox(\xpos,\ypos){\vphantom{#3}#4}}%
\horsize{\tempcounta}{#3}%
\horsize{\tempcountb}{#4}%
\divide \tempcounta by2
\divide \tempcountb by2
\advance \tempcounta by30
\advance \tempcountb by30
\advance \xpos by\tempcounta
\advance \arrowlength by-\tempcounta
\advance \arrowlength by-\tempcountb
\putvector(\xpos,\ypos)(1,0)\arrowlength\arrowtype
\divide \arrowlength by2
\advance \xpos by\arrowlength
\vertsize{\tempcounta}{#5}%
\divide\tempcounta by2
\advance \tempcounta by20
\if a#8 %
   \advance \ypos by\tempcounta
   \putbox(\xpos,\ypos){#5}%
\else
   \advance \ypos by-\tempcounta
   \putbox(\xpos,\ypos){#5}%
\fi}}
 
\def\putvmorphism(#1,#2)[#3`#4`#5]#6#7#8{{%
\xpos #1
\ypos #2
\arrowlength #6
\arrowtype #7
\settowidth{\xlen}{$#5$}%
\putbox(\xpos,\ypos){#3}%
{\advance \ypos by-\arrowlength
\putbox(\xpos,\ypos){#4}}%
{\advance\arrowlength by-140
\advance \ypos by-70
\ifdim\xlen>0pt
   \if m#8%
      \putsplitvector(\xpos,\ypos)\arrowlength\arrowtype
   \else
   \putvector(\xpos,\ypos)(0,-1)\arrowlength\arrowtype
   \fi
\else
   \putvector(\xpos,\ypos)(0,-1)\arrowlength\arrowtype
\fi}%
\ifdim\xlen>0pt
   \divide \arrowlength by2
   \advance\ypos by-\arrowlength
   \if l#8%
      \advance \xpos by-40
      \putrbox(\xpos,\ypos){#5}%
   \else\if r#8%
      \advance \xpos by40
      \putlbox(\xpos,\ypos){#5}%
   \else
      \putbox(\xpos,\ypos){#5}%
   \fi\fi
\fi
}}
 
\def\putsquarep<#1>(#2)[#3;#4`#5`#6`#7]{{%
\setsqparms[#1]%
\setpos(#2)%
\settokens`#3`%
\puthmorphism(\xpos,\ypos)[\tokenc`\tokend`{#7}]{\width}{\arrowtyped}b%
\advance\ypos by \height
\puthmorphism(\xpos,\ypos)[\tokena`\tokenb`{#4}]{\width}{\arrowtypea}a%
\putvmorphism(\xpos,\ypos)[``{#5}]{\height}{\arrowtypeb}l%
\advance\xpos by \width
\putvmorphism(\xpos,\ypos)[``{#6}]{\height}{\arrowtypec}r%
}}
 
\def\putsquare{\@ifnextchar <{\putsquarep}{\putsquarep%
   <\arrowtypea`\arrowtypeb`\arrowtypec`\arrowtyped;\width`\height>}}
\def\square{\@ifnextchar< {\squarep}{\squarep
   <\arrowtypea`\arrowtypeb`\arrowtypec`\arrowtyped;\width`\height>}}
\def\squarep<#1>[#2`#3`#4`#5;#6`#7`#8`#9]{{
\setsqparms[#1]
\diagram
\putsquarep<\arrowtypea`\arrowtypeb`\arrowtypec`
\arrowtyped;\width`\height>
(0,0)[#2`#3`#4`{#5};#6`#7`#8`{#9}]
\enddiagram
}}                                                 
\def\putptrianglep<#1>(#2,#3)[#4`#5`#6;#7`#8`#9]{{%
\settriparms[#1]%
\xpos=#2 \ypos=#3
\advance\ypos by \height
\puthmorphism(\xpos,\ypos)[#4`#5`{#7}]{\height}{\arrowtypea}a%
\putvmorphism(\xpos,\ypos)[`#6`{#8}]{\height}{\arrowtypeb}l%
\advance\xpos by\height
\putmorphism(\xpos,\ypos)(-1,-1)[``{#9}]{\height}{\arrowtypec}r%
}}
 
\def\putptriangle{\@ifnextchar <{\putptrianglep}{\putptrianglep
   <\arrowtypea`\arrowtypeb`\arrowtypec;\height>}}
\def\ptriangle{\@ifnextchar <{\ptrianglep}{\ptrianglep
   <\arrowtypea`\arrowtypeb`\arrowtypec;\height>}}
\def\ptrianglep<#1>[#2`#3`#4;#5`#6`#7]{{
\settriparms[#1]
\diagram
\putptrianglep<\arrowtypea`\arrowtypeb`
\arrowtypec;\height>
(0,0)[#2`#3`#4;#5`#6`{#7}]
\enddiagram
}}                                            
 
\def\putqtrianglep<#1>(#2,#3)[#4`#5`#6;#7`#8`#9]{{%
\settriparms[#1]%
\xpos=#2 \ypos=#3
\advance\ypos by\height
\puthmorphism(\xpos,\ypos)[#4`#5`{#7}]{\height}{\arrowtypea}a%
\putmorphism(\xpos,\ypos)(1,-1)[``{#8}]{\height}{\arrowtypeb}l%
\advance\xpos by\height
\putvmorphism(\xpos,\ypos)[`#6`{#9}]{\height}{\arrowtypec}r%
}}
 
\def\putqtriangle{\@ifnextchar <{\putqtrianglep}{\putqtrianglep
   <\arrowtypea`\arrowtypeb`\arrowtypec;\height>}}
\def\qtriangle{\@ifnextchar <{\qtrianglep}{\qtrianglep
   <\arrowtypea`\arrowtypeb`\arrowtypec;\height>}}
\def\qtrianglep<#1>[#2`#3`#4;#5`#6`#7]{{
\settriparms[#1]
\width=\height                                
\diagram
\putqtrianglep<\arrowtypea`\arrowtypeb`
\arrowtypec;\height>
(0,0)[#2`#3`#4;#5`#6`{#7}]
\enddiagram
}}
 
\def\putdtrianglep<#1>(#2,#3)[#4`#5`#6;#7`#8`#9]{{%
\settriparms[#1]%
\xpos=#2 \ypos=#3
\puthmorphism(\xpos,\ypos)[#5`#6`{#9}]{\height}{\arrowtypec}b%
\advance\xpos by \height \advance\ypos by\height
\putmorphism(\xpos,\ypos)(-1,-1)[``{#7}]{\height}{\arrowtypea}l%
\putvmorphism(\xpos,\ypos)[#4``{#8}]{\height}{\arrowtypeb}r%
}}
 
\def\putdtriangle{\@ifnextchar <{\putdtrianglep}{\putdtrianglep
   <\arrowtypea`\arrowtypeb`\arrowtypec;\height>}}
\def\dtriangle{\@ifnextchar <{\dtrianglep}{\dtrianglep
   <\arrowtypea`\arrowtypeb`\arrowtypec;\height>}}
\def\dtrianglep<#1>[#2`#3`#4;#5`#6`#7]{{
\settriparms[#1]
\width=\height                                
\diagram
\putdtrianglep<\arrowtypea`\arrowtypeb`
\arrowtypec;\height>
(0,0)[#2`#3`#4;#5`#6`{#7}]
\enddiagram
}}
 
\def\putbtrianglep<#1>(#2,#3)[#4`#5`#6;#7`#8`#9]{{%
\settriparms[#1]%
\xpos=#2 \ypos=#3
\puthmorphism(\xpos,\ypos)[#5`#6`{#9}]{\height}{\arrowtypec}b%
\advance\ypos by\height
\putmorphism(\xpos,\ypos)(1,-1)[``{#8}]{\height}{\arrowtypeb}r%
\putvmorphism(\xpos,\ypos)[#4``{#7}]{\height}{\arrowtypea}l%
}}
 
\def\putbtriangle{\@ifnextchar <{\putbtrianglep}{\putbtrianglep
   <\arrowtypea`\arrowtypeb`\arrowtypec;\height>}}
\def\btriangle{\@ifnextchar <{\btrianglep}{\btrianglep
   <\arrowtypea`\arrowtypeb`\arrowtypec;\height>}}
\def\btrianglep<#1>[#2`#3`#4;#5`#6`#7]{{
\settriparms[#1]
\width=\height                               
\diagram
\putbtrianglep<\arrowtypea`\arrowtypeb`
\arrowtypec;\height>
(0,0)[#2`#3`#4;#5`#6`{#7}]
\enddiagram
}}
 
\def\putAtrianglep<#1>(#2,#3)[#4`#5`#6;#7`#8`#9]{{%
\settriparms[#1]%
\xpos=#2 \ypos=#3
{\multiply \height by2
\puthmorphism(\xpos,\ypos)[#5`#6`{#9}]{\height}{\arrowtypec}b}%
\advance\xpos by\height \advance\ypos by\height
\putmorphism(\xpos,\ypos)(-1,-1)[#4``{#7}]{\height}{\arrowtypea}l%
\putmorphism(\xpos,\ypos)(1,-1)[``{#8}]{\height}{\arrowtypeb}r%
}}
 
\def\putAtriangle{\@ifnextchar <{\putAtrianglep}{\putAtrianglep
   <\arrowtypea`\arrowtypeb`\arrowtypec;\height>}}
\def\Atriangle{\@ifnextchar <{\Atrianglep}{\Atrianglep
   <\arrowtypea`\arrowtypeb`\arrowtypec;\height>}}
\def\Atrianglep<#1>[#2`#3`#4;#5`#6`#7]{{
\settriparms[#1]
\width=\height                                     
\diagram
\putAtrianglep<\arrowtypea`\arrowtypeb`
\arrowtypec;\height>
(0,0)[#2`#3`#4;#5`#6`{#7}]
\enddiagram
}}
 
\def\putAtrianglepairp<#1>(#2)[#3;#4`#5`#6`#7`#8]{{%
\settripairparms[#1]%
\setpos(#2)%
\settokens`#3`%
\puthmorphism(\xpos,\ypos)[\tokenb`\tokenc`{#7}]{\height}{\arrowtyped}b%
\advance\xpos by\height
\puthmorphism(\xpos,\ypos)[\phantom{\tokenc}`\tokend`{#8}]%
{\height}{\arrowtypee}b%
\advance\ypos by\height
\putmorphism(\xpos,\ypos)(-1,-1)[\tokena``{#4}]{\height}{\arrowtypea}l%
\putvmorphism(\xpos,\ypos)[``{#5}]{\height}{\arrowtypeb}m%
\putmorphism(\xpos,\ypos)(1,-1)[``{#6}]{\height}{\arrowtypec}r%
}}
 
\def\putAtrianglepair{\@ifnextchar <{\putAtrianglepairp}{\putAtrianglepairp%
   <\arrowtypea`\arrowtypeb`\arrowtypec`\arrowtyped`\arrowtypee;\height>}}
\def\Atrianglepair{\@ifnextchar <{\Atrianglepairp}{\Atrianglepairp%
   <\arrowtypea`\arrowtypeb`\arrowtypec`\arrowtyped`\arrowtypee;\height>}}
 
\def\Atrianglepairp<#1>[#2;#3`#4`#5`#6`#7]{{
\settripairparms[#1]
\settokens`#2`
\width=\height                                
\diagram
\putAtrianglepairp                            
<\arrowtypea`\arrowtypeb`\arrowtypec`
\arrowtyped`\arrowtypee;\height>
(0,0)[{#2};#3`#4`#5`#6`{#7}]
\enddiagram
}}
 
\def\putVtrianglep<#1>(#2,#3)[#4`#5`#6;#7`#8`#9]{{%
\settriparms[#1]%
\xpos=#2 \ypos=#3
\advance\ypos by\height
{\multiply\height by2
\puthmorphism(\xpos,\ypos)[#4`#5`{#7}]{\height}{\arrowtypea}a}%
\putmorphism(\xpos,\ypos)(1,-1)[`#6`{#8}]{\height}{\arrowtypeb}l%
\advance\xpos by\height
\advance\xpos by\height
\putmorphism(\xpos,\ypos)(-1,-1)[``{#9}]{\height}{\arrowtypec}r%
}}
 
\def\putVtriangle{\@ifnextchar <{\putVtrianglep}{\putVtrianglep
   <\arrowtypea`\arrowtypeb`\arrowtypec;\height>}}
\def\Vtriangle{\@ifnextchar <{\Vtrianglep}{\Vtrianglep
   <\arrowtypea`\arrowtypeb`\arrowtypec;\height>}}
\def\Vtrianglep<#1>[#2`#3`#4;#5`#6`#7]{{
\settriparms[#1]
\width=\height                                 
\diagram
\putVtrianglep<\arrowtypea`\arrowtypeb`
\arrowtypec;\height>
(0,0)[#2`#3`#4;#5`#6`{#7}]
\enddiagram
}}
 
\def\putVtrianglepairp<#1>(#2)[#3;#4`#5`#6`#7`#8]{{
\settripairparms[#1]%
\setpos(#2)%
\settokens`#3`%
\advance\ypos by\height
\putmorphism(\xpos,\ypos)(1,-1)[`\tokend`{#6}]{\height}{\arrowtypec}l%
\puthmorphism(\xpos,\ypos)[\tokena`\tokenb`{#4}]{\height}{\arrowtypea}a%
\advance\xpos by\height
\puthmorphism(\xpos,\ypos)[\phantom{\tokenb}`\tokenc`{#5}]%
{\height}{\arrowtypeb}a%
\putvmorphism(\xpos,\ypos)[``{#7}]{\height}{\arrowtyped}m%
\advance\xpos by\height
\putmorphism(\xpos,\ypos)(-1,-1)[``{#8}]{\height}{\arrowtypee}r%
}}
 
\def\putVtrianglepair{\@ifnextchar <{\putVtrianglepairp}{\putVtrianglepairp%
    <\arrowtypea`\arrowtypeb`\arrowtypec`\arrowtyped`\arrowtypee;\height>}}
\def\Vtrianglepair{\@ifnextchar <{\Vtrianglepairp}{\Vtrianglepairp%
    <\arrowtypea`\arrowtypeb`\arrowtypec`\arrowtyped`\arrowtypee;\height>}}
\def\Vtrianglepairp<#1>[#2;#3`#4`#5`#6`#7]{{
\settripairparms[#1]
\settokens`#2`
\diagram
\putVtrianglepairp                             
<\arrowtypea`\arrowtypeb`\arrowtypec`
\arrowtyped`\arrowtypee;\height>
(0,0)[{#2};#3`#4`#5`#6`{#7}]
\enddiagram
}}

\def\putCtrianglep<#1>(#2,#3)[#4`#5`#6;#7`#8`#9]{{%
\settriparms[#1]%
\xpos=#2 \ypos=#3
\advance\ypos by\height
\putmorphism(\xpos,\ypos)(1,-1)[``{#9}]{\height}{\arrowtypec}l%
\advance\xpos by\height
\advance\ypos by\height
\putmorphism(\xpos,\ypos)(-1,-1)[#4`#5`{#7}]{\height}{\arrowtypea}l%
{\multiply\height by 2
\putvmorphism(\xpos,\ypos)[`#6`{#8}]{\height}{\arrowtypeb}r}%
}}
 
\def\putCtriangle{\@ifnextchar <{\putCtrianglep}{\putCtrianglep
    <\arrowtypea`\arrowtypeb`\arrowtypec;\height>}}
\def\Ctriangle{\@ifnextchar <{\Ctrianglep}{\Ctrianglep
    <\arrowtypea`\arrowtypeb`\arrowtypec;\height>}}
\def\Ctrianglep<#1>[#2`#3`#4;#5`#6`#7]{{
\settriparms[#1]
\width=\height                               
\diagram
\putCtrianglep<\arrowtypea`\arrowtypeb`
\arrowtypec;\height>
(0,0)[#2`#3`#4;#5`#6`{#7}]
\enddiagram
}}                                           
\def\putDtrianglep<#1>(#2,#3)[#4`#5`#6;#7`#8`#9]{{%
\settriparms[#1]%
\xpos=#2 \ypos=#3
\advance\xpos by\height \advance\ypos by\height
\putmorphism(\xpos,\ypos)(-1,-1)[``{#9}]{\height}{\arrowtypec}r%
\advance\xpos by-\height \advance\ypos by\height
\putmorphism(\xpos,\ypos)(1,-1)[`#5`{#8}]{\height}{\arrowtypeb}r%
{\multiply\height by 2
\putvmorphism(\xpos,\ypos)[#4`#6`{#7}]{\height}{\arrowtypea}l}%
}}
 
\def\putDtriangle{\@ifnextchar <{\putDtrianglep}{\putDtrianglep
    <\arrowtypea`\arrowtypeb`\arrowtypec;\height>}}
\def\Dtriangle{\@ifnextchar <{\Dtrianglep}{\Dtrianglep
   <\arrowtypea`\arrowtypeb`\arrowtypec;\height>}}
\def\Dtrianglep<#1>[#2`#3`#4;#5`#6`#7]{{
\settriparms[#1]
\width=\height                              
\diagram
\putDtrianglep<\arrowtypea`\arrowtypeb`
\arrowtypec;\height>
(0,0)[#2`#3`#4;#5`#6`{#7}]
\enddiagram
}}                                          
\def\setrecparms[#1`#2]{\width=#1 \height=#2}%
 
\def\recursep<#1`#2>[#3;#4`#5`#6`#7`#8]{{\m@th
\width=#1 \height=#2
\settokens`#3`
\settowidth{\tempdimen}{$\tokena$}
\ifdim\tempdimen=0pt
  \savebox{\tempboxa}{\hbox{$\tokenb$}}%
  \savebox{\tempboxb}{\hbox{$\tokend$}}%
  \savebox{\tempboxc}{\hbox{$#6$}}%
\else
  \savebox{\tempboxa}{\hbox{$\hbox{$\tokena$}\times\hbox{$\tokenb$}$}}%
  \savebox{\tempboxb}{\hbox{$\hbox{$\tokena$}\times\hbox{$\tokend$}$}}%
  \savebox{\tempboxc}{\hbox{$\hbox{$\tokena$}\times\hbox{$#6$}$}}%
\fi
\ypos=\height
\divide\ypos by 2
\xpos=\ypos
\advance\xpos by \width
\bfig
\putCtrianglep<-1`1`1;\ypos>(0,0)[`\tokenc`;#5`#6`{#7}]%
\puthmorphism(\ypos,0)[\tokend`\usebox{\tempboxb}`{#8}]{\width}{-1}b%
\puthmorphism(\ypos,\height)[\tokenb`\usebox{\tempboxa}`{#4}]{\width}{-1}a%
\advance\ypos by \width
\putvmorphism(\ypos,\height)[``\usebox{\tempboxc}]{\height}1r%
\efig
}}
 
\def\recurse{\@ifnextchar <{\recursep}{\recursep<\width`\height>}}
 
\def\puttwohmorphisms(#1,#2)[#3`#4;#5`#6]#7#8#9{{%
\puthmorphism(#1,#2)[#3`#4`]{#7}0a
\ypos=#2
\advance\ypos by 20
\puthmorphism(#1,\ypos)[\phantom{#3}`\phantom{#4}`#5]{#7}{#8}a
\advance\ypos by -40
\puthmorphism(#1,\ypos)[\phantom{#3}`\phantom{#4}`#6]{#7}{#9}b
}}
 
\def\puttwovmorphisms(#1,#2)[#3`#4;#5`#6]#7#8#9{{%
\putvmorphism(#1,#2)[#3`#4`]{#7}0a
\xpos=#1
\advance\xpos by -20
\putvmorphism(\xpos,#2)[\phantom{#3}`\phantom{#4}`#5]{#7}{#8}l
\advance\xpos by 40
\putvmorphism(\xpos,#2)[\phantom{#3}`\phantom{#4}`#6]{#7}{#9}r
}}
 
\def\puthcoequalizer(#1)[#2`#3`#4;#5`#6`#7]#8#9{{%
\setpos(#1)%
\puttwohmorphisms(\xpos,\ypos)[#2`#3;#5`#6]{#8}11%
\advance\xpos by #8
\puthmorphism(\xpos,\ypos)[\phantom{#3}`#4`#7]{#8}1{#9}
}}
 
\def\putvcoequalizer(#1)[#2`#3`#4;#5`#6`#7]#8#9{{%
\setpos(#1)%
\puttwovmorphisms(\xpos,\ypos)[#2`#3;#5`#6]{#8}11%
\advance\ypos by -#8
\putvmorphism(\xpos,\ypos)[\phantom{#3}`#4`#7]{#8}1{#9}
}}
 
\def\putthreehmorphisms(#1)[#2`#3;#4`#5`#6]#7(#8)#9{{%
\setpos(#1) \settypes(#8)
\if a#9 %
     \vertsize{\tempcounta}{#5}%
     \vertsize{\tempcountb}{#6}%
     \ifnum \tempcounta<\tempcountb \tempcounta=\tempcountb \fi
\else
     \vertsize{\tempcounta}{#4}%
     \vertsize{\tempcountb}{#5}%
     \ifnum \tempcounta<\tempcountb \tempcounta=\tempcountb \fi
\fi
\advance \tempcounta by 60
\puthmorphism(\xpos,\ypos)[#2`#3`#5]{#7}{\arrowtypeb}{#9}
\advance\ypos by \tempcounta
\puthmorphism(\xpos,\ypos)[\phantom{#2}`\phantom{#3}`#4]{#7}{\arrowtypea}{#9}
\advance\ypos by -\tempcounta \advance\ypos by -\tempcounta
\puthmorphism(\xpos,\ypos)[\phantom{#2}`\phantom{#3}`#6]{#7}{\arrowtypec}{#9}
}}
 
\def\setarrowtoks[#1`#2`#3`#4`#5`#6]{%
\def\toka{#1}
\def\tokb{#2}
\def\tokc{#3}
\def\tokd{#4}
\def\toke{#5}
\def\tokf{#6}
}
\def\hex{\@ifnextchar <{\hexp}{\hexp<1000`400>}}
\def\hexp<#1`#2>[#3`#4`#5`#6`#7`#8;#9]{%
\setarrowtoks[#9]
\yext=#2 \advance \yext by #2
\xext=#1 \advance\xext by \yext
\bfig
\putCtriangle<-1`0`1;#2>(0,0)[`#5`;\tokb``\tokd]
\xext=#1 \yext=#2 \advance \yext by #2
\putsquare<1`0`0`1;\xext`\yext>(#2,0)[#3`#4`#7`#8;\toka```\tokf]
\advance \xext by #2
\putDtriangle<0`1`-1;#2>(\xext,0)[`#6`;`\tokc`\toke]
\efig
}
\makeatother

\let\ssection=\section
\renewcommand{\section}{\setcounter{equation}{0}\ssection}

\newtheorem{definition}{Definition}[section]
\newtheorem{theorem}{Theorem}[section]

\newenvironment{proof}[1][Proof]{\noindent\textbf{#1.} }{\ \rule{0.5em}{0.5em}}

\newcommand\mathC{\mkern1mu\raise2.2pt\hbox{$\scriptscriptstyle|$}
        {\mkern-7mu\rm C}}            
\newcommand{\mathR}{{\rm I\! R}}      

\newcommand\ie{{i.e.},}

\newcommand{\Ga}{\Gamma}
\newcommand{\ka}{\kappa}
\renewcommand\l{\lambda}
\newcommand\s{\sigma}
\newcommand\Si{\Sigma}
\newcommand\De{\Delta}

\renewcommand{\O}{\Omega}

\newcommand{\map}{\rightarrow}               

\newcommand{\sa}{{\rm sa}}

\newcommand{\piqt}{\pi_{{\rm qt}}}         

\newcommand{\A}{{\hat A}}
\newcommand{\U}{{\hat U}}
\renewcommand{\P}{{\hat P}}
\renewcommand{\S}{{\cal S}}
\newcommand{\Hi}{{\cal H}}

\newcommand\BH{\mathcal{B(H)}}
\newcommand\PH{\mathcal{P(H)}}
\newcommand\PV{\mathcal{P}(V)}
\newcommand\UH{\mathcal{U(H)}}
\newcommand\BV{{\rm BV}(\Ob{\V{}},\mathR)}

\newcommand\TO{\mathbb{T}}   

\newcommand\bra[1]{\langle #1|\,}
\newcommand\ket[1]{\,|#1\rangle}
\newcommand\eq[1]{(\ref{#1})}
\newcommand\eqs[2]{(\ref{#1}--\ref{#2})}
\newcommand\SAin[1]{\mbox{``}A\,\varepsilon\,#1\mbox{''}}
\newcommand\Ain[1]{A\,\varepsilon\,#1}

\newcommand\daso{\delta^o}
\newcommand\dasi{\delta^i}
\newcommand\dasoB[1]{\delta^o(\hat{#1})}        

\newcommand\dasiB[1]{\delta^i(\hat{#1})}        
\newcommand\dastoo[2]{\delta^o(\hat{#2})_{#1}}  
\newcommand\dastoi[2]{\delta^i(\hat{#2})_{#1}}  

\newcommand\dasB[1]{\breve{\delta}(#1)}
\newcommand\dasBV[2]{\breve{\delta}(#2)_{#1}}
\newcommand\dasBi[1]{\breve{\delta}^i(#1)}
\newcommand\dasBVi[2]{\breve{\delta}^i(#2)_{#1}}
\newcommand\dasBo[1]{\breve{\delta}^o(#1)}
\newcommand\dasBVo[2]{\breve{\delta}^o(#2)_{#1}}

\renewcommand\L[1]{\mathcal{L}({#1})}
\newcommand\PL[1]{{\cal PL}(#1)}
\renewcommand\sp[1]{{\rm sp}(\hat A)}
\newcommand\GT[1]{\overline {#1}}

\newcommand\TVal[2]{\nu\big(#1;#2\big)}         
\newcommand\Hom[3]{{\rm Hom}_{#1}\big(#2,#3\big)}

\newcommand\ps[1]{\underline{#1}}        
\newcommand{\G}{\ps{O}}                   
\renewcommand{\H}{\ps{I}}                 
\newcommand{\dG}{\ps{\mkern1mu\raise2.5pt\hbox{$\scriptscriptstyle|$}
        {\mkern-7mu\rm O}}}                 
\newcommand{\dH}{\ps{{\rm I\! I}}}        
\newcommand{\dOU}{\ps{\mkern1mu\raise2.5pt\hbox{$\scriptscriptstyle|$}
        {\mkern-7mu\rm U}}}               

\newcommand{\Sig}{\ps{\Sigma}}            
\newcommand{\PSig}{P_{{\rm cl}}\Sig}      
\newcommand{\R}{{\cal R}}                 

\newcommand{\SpA}{\ps{\sp{A}^{\succeq}}}
\newcommand{\SR}{\ps{{\mathR}^\succeq}}         
\newcommand{\OP}{\ps{{\mathR}^\preceq}}         
\newcommand{\PR}[1]{\ps{#1^\leftrightarrow}}

\newcommand{\kSR}{k(\SR)}                   

\newcommand\Ob[1]{{\rm Ob(#1)}}
\newcommand\Subcl[1]{{\rm Sub}_{{\rm cl}}(#1)} 

\newcommand\Set{{\bf Sets}}                    
\newcommand\SetH[1]{\Set^{{\V{#1}}^{\rm op}}}  

\newcommand\V[1]{{\cal V}(\Hi_{#1})}           

\begin{document}

\begin{titlepage}

\begin{center}
{\large\bf A Topos Foundation for Theories of Physics:

III.\ The Representation of Physical Quantities With Arrows
$\dasBo{A}:\Sig\map\SR$}
\end{center}

\vspace{0.8 truecm}
\begin{center}
        A.~D\"oring\footnote{email: a.doering@imperial.ac.uk}\\[10pt]

\begin{center}                      and
\end{center}

        C.J.~Isham\footnote{email: c.isham@imperial.ac.uk}\\[10pt]

        The Blackett Laboratory\\ Imperial College of Science,
        Technology \& Medicine\\ South Kensington\\ London SW7 2BZ\\
\end{center}

\begin{center}
6 March, 2007
\end{center}

\begin{abstract}
This paper is the third in a series whose goal is to develop a
fundamentally new way of viewing theories of physics. Our basic
contention is that constructing a theory of physics is equivalent
to finding a representation in a topos of a certain formal
language that is attached to the system.

In paper II, we studied the topos representations of the
propositional language $\PL{S}$ for the case of quantum theory,
and in the present paper we do the same thing for the, more
extensive, local language $\L{S}$. One of the main achievements is
to find  a topos representation for self-adjoint operators. This
involves showing that, for any physical quantity $A$, there is an
arrow $\dasBo{A}:\Sig\map\SR$, where $\SR$ is the quantity-value
object for this theory. The construction of $\dasBo{\A}$ is an
extension of the daseinisation of projection operators that was
discussed in paper II.

The object $\SR$ is  a monoid-object only in the topos,
$\tau_\phi=\SetH{}$, of the theory, and to enhance the
applicability of the formalism, we apply to $\SR$ a topos analogue
of the Grothendieck extension of a monoid to a group. The
resulting object, $\kSR$, is an abelian group-object in
$\tau_\phi$. We also discuss another candidate, $\PR{\mathR}$, for
the quantity-value object. In this presheaf, both inner and
outer daseinisation are used in a symmetric way.

Finally, there is a brief discussion of the role of unitary
operators in the quantum topos scheme.
\end{abstract}
\end{titlepage}

\section{Introduction}
This is the third in a series of papers whose aim is to construct
a general framework within which theories of physics can be
developed in a topos other than that of sets. A theme that runs
throughout this work is  the idea that any `non-classical' theory,
for example, quantum theory, can be presented in a way that `looks
like' classical physics, except that the topos is generally not
the topos of sets. This approach provides a new set of tools with
which to construct theories of physics. At the conceptual level,
the analogue with classical physics leads to the scheme  being
`neo-realist'.

In paper I, we introduced the idea of associating a formal
language with each physical system $S$ \cite{DI(1)}. Constructing
a theory of physics is then equivalent to finding a representation
of this language in a topos. Two different kinds of language are
discussed in paper I:  a simple propositional language $\PL{S}$;
and a more sophisticated, higher-order language (a `local'
language) $\L{S}$.

The language $\PL{S}$ provides a deductive system (using
intuitionistic logic) and hence provides a way of making
statements about the system $S$. However, a purely propositional
language is limited in scope: at the very least, one would like to
have a `first-order' language, so that the phrases `for all' and
`there exists' can be used.

But such a language is still rudimentary in so far as many
features of a physical theory would lie \emph{outside} its scope,
and are introduced only when constructing a representation. For
example, in classical mechanics, the entities that lie outside the
language are (i) the state space $\S$; (ii) the choice of $\mathR$
as the set in which physical quantities take their values; (iii)
the specific subset $\De\subseteq\mathR$ that is used in the
proposition $\SAin\De$ and (iv) the real-valued functions on $\S$
that represent physical quantities.

For this reason, the next step in paper I was to assign to each
physical system $S$, a  more powerful, typed language, $\L{S}$.
Our general scheme can then be understood as the task of finding
representations of $\L{S}$ in various topoi. The language $\L{S}$
has two `ground-type' symbols, $\Si$ and $\R$, and a set of
`function symbols', written rather suggestively as the string of
characters `$A:\Si\map\R$'. These are the linguistic precursors
of, respectively,  the state object, the quantity-value object,
and the arrows between them that represent physical quantities. A
symbol $\tilde{\De}$ can be introduced as a variable of type
$P\R$. By these means, the entities that lie outside the
propositional language $\PL{S}$ are all brought `inside' the local
language $\L{S}$.

The second paper in the series dealt with the topos representation
of the propositional language $\PL{S}$ in quantum theory
\cite{DI(2)}. In a sense this paper is a side-line to our main
programme which is concerned with the local language $\L{S}$. In
fact, logically speaking, we could have omitted our study of the
representations of $\PL{S}$. However, the $\PL{S}$ material links
most closely with the original work on the use of topos ideas in
quantum theory. Also, as we will see, the vital concept of
`daseinisation' of propositions has a natural extension to
physical quantities in general, and this plays a key role in the
remaining papers in the series.

\pagebreak

In the present paper, we return to the language $\L{S}$ and seek a
topos representation, $\phi$, for quantum theory. As discussed in
paper II, the topos of the $\PL{S}$-representation  is $\SetH{}$:
the topos of presheaves over the category, $\V{}$, of unital,
abelian subalgebras of the algebra $\BH$ of all bounded operators
on the quantum Hilbert space $\Hi$. We shall use the same topos
for the $\L{S}$-representation, with  the spectral presheaf,
$\Sig$, being identified as the $\SetH{}$-representative,
$\Si_\phi$, of the ground-type symbol $\Si$. Thus $\Sig$ is the
state object, and, therefore, propositions are represented by
sub-objects of $\Sig$ ; just as, in classical physics, a
proposition about the system is represented by a subset of the
classical state space.

The steps in finding the representation of $\L{S}$ are first to
identify the quantity-value object, $\R_\phi$; and then to show
how to represent a physical quantity, $A$, by an arrow
$\dasB{A}:\Sig\map\R_\phi$. Both  problems are solved in the
present paper.

The plan of the paper is as follows. We start in Section
\ref{Sec:deG} by defining the inner and outer daseinisations of an
arbitrary self-adjoint operator $\A\in\BH$. These constructions
utilise the daseinisations of projection operators that were
discussed in paper II.  The result is two quantities, $\dasiB{A}$
and $\dasoB{A}$, that can be identified as  global elements of two
new presheaves: the `inner-' and the `outer de Groote presheaf',
respectively.

In Section \ref{Sec:psSR}, for each stage $V\in\Ob{\V{}}$, we take
the Gel'fand transforms of $\dastoo{V}{A}$,  and show how the
target spaces of these transforms fit together to give a new
presheaf, $\SpA$, which is a topos extension of the spectrum,
$\sp{A}$, of $\A$. For any self-adjoint operator $\A$, the target
presheaf $\SpA$ can be embedded as a sub-object of a single
presheaf $\SR$. In this way, $\SR$ is provisionally identified as
the  quantity-value presheaf for quantum theory in the topos
$\SetH{}$. There is also an isomorphic presheaf, $\OP$, that is
constructed using inner daseinisation; and a third presheaf,
$\PR{\mathR}$, that combines both inner and outer daseinisation.
This is the presheaf that gives the most natural physical
interpretation of the arrows $\dasB{A}:\Sig\map\PR{\mathR}$.

However, algebraically speaking, $\SR$, $\OP$, and $\PR{\mathR}$
are only additive monoid-objects in the topos, whereas the set of
real numbers, $\mathR$, employed in standard physics, is an
abelian group (in fact, even an abelian ring). This limits what
can be done with these presheaves and motivates us to apply a
standard trick in algebraic topology. This is the `Grothendieck
completion' of a monoid to give a group, which we adapt to produce
a presheaf, $\kSR$, that extends the monoid-object $\SR$ to a full
group-object in the topos $\SetH{}$. We show that the square can
be taken of certain elements in $\kSR$, which enables us to define
a kind of dispersion for daseinised self-adjoint operators. We
also show that the presheaf $\kSR$ is closely related to
$\PR{\mathR}$.

Finally, in Section \ref{Sec:Unitary}, we recall the role of
unitary operators in standard quantum theory, and show how each
such operator, $\U$, can be made to act as a functor from the
topos $\SetH{}$ to itself. This leads to the topos analogue of the
covariance statements that are associated with unitary operators
in quantum theory.

\section{The de Groote Presheaves of Physical Quantities}
\label{Sec:deG}
\subsection{Background Remarks}
Our task is to consider the representation of the local language,
$\L{S}$, in the case of quantum theory. We assume that the
relevant topos  is the same as that used for the propositional
language $\PL{S}$, \ie\ $\SetH{}$, but the emphasis is very
different.

From a physics perspective, the key symbols in $\L{S}$ are the
ground-type symbols, $\Si$ and $\mathcal R$---the linguistic
precursors of the state object and the quantity-value object
respectively---and the function symbols $A:\Si\map\R$, which are
the precursors of physical quantities. In the quantum-theory
representation, $\phi$, of $\L{S}$,  the representation,
$\Si_\phi$, of $\Si$ is defined to be the spectral presheaf $\Sig$
in the topos  $\SetH{}$.

The critical question is to find the object, $\R_\phi$
(provisionally denoted as a presheaf $\ps{\R}$), in $\SetH{}$ that
represents $\R$, and is hence the quantity-value object. One might
think that $\ps{\R}$ is the real-number object in the topos
$\SetH{}$, but that is wrong, and the right answer cannot just be
guessed. In fact, the correct choice for $\ps{\R}$ is found
indirectly by considering a related question: namely, how to
represent each function symbol $A:\Si\map\R$, with a concrete
arrow $A_\phi:\Si_\phi\map\R_\phi$ in  $\SetH{}$, \ie\ with a
natural transformation $\breve{A}:\Sig\map\ps{\R}$ between the
presheaves $\Sig$ and $\ps{\R}$.

\subsection{The Outer and Inner Presheaves}
The daseinisation operations on projection operators were
introduced by de Groote \cite{deG05}, and exploited by us in
\cite{DI(2)}. They are defined as follows. {\definition If $\hat
P$ is a projection operator, and $V\in\Ob{\V{}}$ is any
context/stage,  we define:
\begin{enumerate}
\item The `outer daseinisation' operation is
\begin{equation}
        \dastoo{V}{P}:=\bigwedge\big\{\hat{Q}\in\PV\mid
                \hat{Q}\succeq \P\big\}.\label{Def:dasouter}
\end{equation}
where `$\,\succeq$' denotes the usual ordering of projection
operators, and where $\PV$ is the set of all projection operators
in $V$.

\item Similarly, the `inner daseinisation' operation is defined in
the context $V$ as
\begin{equation}
        \dastoi{V}{P}:=\bigvee\big\{\hat{Q}\in\PV\mid
                \hat{Q}\preceq \P\big\}. \label{Def:dasinner}
\end{equation}
\end{enumerate}
} \noindent Thus $\dastoo{V}{P}$ is the best approximation to
$\hat P$ in $V$ from `above', being the smallest projection in $V$
that is larger than or equal to $\hat P$. Similarly,
$\dastoi{V}{P}$ is the best approximation to $\hat P$ from
`below', being the largest projection in $V$ that is smaller than
or equal to $\hat P$.

Daseinisation was used in paper II in the construction of the
`outer' and `inner' presheaves $\G$ and $\H$, which are defined as
follows: {\definition The outer presheaf, $\G$,   is defined over
the category $\V{}$ by:
\begin{enumerate}
\item[(i)] On objects $V\in\Ob{\V{}}$:  $\G_V:=\PV$.

\item[(ii)] On morphisms $i_{V^{\prime}V}:V^{\prime }\subseteq V:$
Define $\G(i_{V^{\prime} V}):\G_V \map\G_{V^{\prime}}$ by
$\G(i_{V^{\prime}V})(\hat{\alpha}):=\dastoo{V^\prime}{\alpha}$ for
all $\hat\alpha\in\PV$.
 \end{enumerate}
}\noindent Evidently, the map $V\mapsto\dastoo{V}{P}$ defines a
global element, $\dasoB{P}$, of the presheaf $\G$.

{\definition The inner presheaf, $\H$, is defined over the
category $\V{}$ by:
\begin{enumerate}
\item[(i)] On objects $V\in\Ob{\V{}}$: $\H_V:=\PV$.

\item[(ii)] On morphisms $i_{V^{\prime}V}:V^{\prime }\subseteq V$:
Define $\H(i_{V^{\prime}\, V}):\H_V \map\H_{V^{\prime}}$ by
$\H(i_{V^{\prime}V})(\hat{\alpha}):=\dastoi{V^\prime}{\alpha}$ for
all $\hat\alpha\in\PV$.
\end{enumerate}
} \noindent Evidently, the map $V\mapsto\dastoi{V}{P}$ defines a
global element, $\dasiB{P}$, of the inner presheaf $\H$.

In paper II, we showed that the outer presheaf is a sub-object of
the power object $\PSig$ (in the category $\SetH{}$), and hence
that the global element $\dasoB{P}$ of $\G$ determines a (clopen)
sub-object of the spectral presheaf $\Sig$. By these means, the
quantum logic of the lattice $\PH$ is mapped into the Heyting
algebra of the set, $\Subcl{\Sig}$, of clopen sub-objects of
$\Sig$.

Our task in the present paper is to perform the second stage of
the quantum programme: namely  (i) identify the quantity-value
presheaf, $\ps\R$; and (ii) show that any physical quantity can be
represented by an arrow from $\Sig$ to $\ps\R$.

\subsection{The Daseinisation of an Arbitrary Self-Adjoint Operator}
\subsubsection{Spectral Families and Spectral Order}
We now want to extend the daseinisation operations from
projections to arbitrary (bounded) self-adjoint operators. To this
end, consider first a bounded, self-adjoint operator, $\A$, whose
spectrum is purely discrete. Then the spectral theorem can be used
to write $\A=\sum_{i=1}^\infty a_i\hat P_i$ where $a_1,a_2,\ldots
$ are the eigenvalues of $\A$, and $\hat P_1,\hat P_2,\ldots $ are
the spectral projection operators onto the corresponding
eigenspaces.

A construction that comes immediately to mind is to use the
daseinisation operation on projections to define
\begin{equation}
\delta^o(\A)_V:= \sum_{i=1}^\infty a_i\,\dasoB{P_i}_V
                                                \label{Def:DasAWrong!}
\end{equation}
for each stage $V$. However, this procedure is rather unnatural.
For one thing, the  projections, $\P_i$, $i=1,2,\ldots$ form a
complete orthonormal set:
\begin{eqnarray}
        \sum_{i=1}^\infty \P_i&=&\hat 1,\\
        \P_i\P_j&=&\delta_{ij}\P_i,
\end{eqnarray}
whereas, in general, the collection of daseinised projections,
$\dasoB{P_i}_V$, $1=1,2,\ldots$ will not satisfy either of these
conditions. In addition, it is hard to see how the expression
$\dasoB{A}_V:= \sum_{i=1}^\infty a_i\,\dasoB{P_i}_V$ can be
generalised to operators, $\A$, with a continuous spectrum.

The answer to this conundrum lies in the work of de Groote. He
realised that although it is not  useful to daseinise the spectral
projections of an operator $\A$, it \emph{is} possible to
daseinise the \emph{spectral family} of $\A$ \cite{deG05}.

\paragraph{Spectral families.}
We first recall that a {\em spectral family} is a family of
projection operators $\hat E_\l$, $\l\in\mathR$, with the
following properties:
\begin{enumerate}
\item If $\l_2\leq\l _1$ then $\hat E_{\l_2}\preceq\hat E_{\l_1}$.
\item The net $\l\mapsto \hat E_\l$ of projection operators in the
lattice $\PH$  is bounded above by $\hat 1$, and below by $\hat
0$. In fact,
\begin{eqnarray}
                \lim_{\l\map\infty} \hat E_\l &=&\hat 1,\\
                \lim_{\l\map -\infty}\hat E_\l&=&\hat 0.
\end{eqnarray}
\item The map $\l\mapsto \hat E_\l$ is right continuous:\footnote{
It is a matter of convention whether one chooses right-continuous
or left-continuous.}
\begin{equation}
  \bigwedge_{\epsilon\downarrow 0} \hat E_{\l+\epsilon}=\hat E_\l
\end{equation}
for all $\l\in\mathR$.
\end{enumerate}
The spectral theorem asserts that for any self-adjoint operator
$\A$, there exists a spectral family, $\l\mapsto \hat E^A_{\l}$,
such that
\begin{equation}
        \A=\int_\mathR \l\, d \hat E^A_{\l}   \label{SpTh}
\end{equation}
We are only concerned with bounded operators, and so the (weak
Stieljes) integral in \eq{SpTh} is really over the bounded
spectrum of $\A$ which, of course, is a compact subset of
$\mathR$. Conversely, given a bounded spectral family $\{\hat
E_\l\}_{\l\in\mathR}$,\footnote{I.e., there are $a,b\in\mathR$
such that $\hat E_\l=\hat 0$ for all $\l\leq a$ and $\hat
E_\l=\hat 1$ for all $\l\geq b$.} there is a bounded self-adjoint
operator $\A$ such that $\A=\int_\mathR \l\, d \hat E_{\l}$.

\paragraph{The spectral order.}
A key element for our work is the so-called {\em spectral order}
that was introduced in \cite{Ols71}.\footnote{The spectral order
was later reinvented by de Groote, see \cite{deG04}.} It is
defined as follows.  Let $\A$ and $\hat B$ be (bounded)
self-adjoint operators with spectral families $\{\hat
E^A_\l\}_{\l\in\mathR}$ and $\{\hat E^B_\l\}_{\l\in\mathR}$,
respectively. Then define:
\begin{equation}
        \A\preceq_s\hat B\mbox{ if and only if }
        \hat E^B_\l\preceq\hat E^A_\l \mbox{ for all $\l\in\mathR$}.
        \label{Def:ApreceqsB}
\end{equation}
It is easy to see that \eq{Def:ApreceqsB} defines a genuine
partial ordering on $\BH_\sa$ (the self-adjoint operators in
$\BH$). In fact, $\BH_\sa$ is a `boundedly complete' lattice with
respect to the spectral order, \ie\ each bounded set $S$ of
self-adjoint operators has a minimum $\bigwedge S\in\BH_\sa$ and a
maximum $\bigvee S\in\BH_\sa$ with respect to this order.

If $\hat P,\hat Q$ are projections, then
\begin{equation}
        \hat P\preceq_s\hat Q \mbox{ if and only if }
        \hat P\preceq\hat Q,             \label{[PQ]=0->s=}
\end{equation}
so the spectral order coincides with the usual partial order on
$\PH$. To ensure this, the `reverse' relation in
\eq{Def:ApreceqsB} is necessary, since the spectral family of a
projection $\hat P$ is given by
\begin{equation}
E_{\l}^{\hat P}=\left\{
\begin{tabular}
[c]{ll}%
$\hat{0}$ & if $\l<0$\\
$\hat{1}-\P$ & if $0\leq\l<1$\\
$\hat{1}$ & if $\l\geq1.$%
\end{tabular}
\right.
\end{equation}

If $\A,\hat B$ are self-adjoint operators such that (i) either
$\A$ or $\hat B$ is a projection, or (ii) $[\A,\hat B]=\hat 0$,
then $\A\preceq_s\hat B \mbox{ if and only if } \A\preceq\hat B$.
Here `$\preceq$' denotes the usual ordering on
$\BH_\sa$.\footnote{ The `usual' ordering is $\A\preceq\hat B$ if
$\bra\psi\A\ket\psi \leq \bra\psi\hat B\ket\psi$ for all vectors
$\ket\psi\in\mathcal H$.}

Moreover, if $\A,\hat B$ are arbitrary self-adjoint operators,
then $\A\preceq_s\hat B$ implies $\A\preceq\hat B$, but not vice
versa in general. Thus the spectral order is a partial order on
$\BH_\sa$ that is coarser than the usual one.

\subsubsection{Daseinisation of Self-Adjoint Operators.}
De Groote's crucial observation was the following. Let
$\l\mapsto\hat E_\l$ be a spectral family in $\PH$ (or,
equivalently, a self-adjoint operator $\A$). Then, for each stage
$V$, the following maps:
\begin{eqnarray}
\l&\mapsto& \bigwedge_{\mu>\l}\dastoo{V}{E_\mu} \label{dasoE}\\
\l&\mapsto& \dastoi{V}{E_\l}                    \label{dasiE}
\end{eqnarray}
also define spectral families.\footnote{The reason  \eq{dasoE} and
\eq{dasiE} have a different form is that
$\l\mapsto\dastoi{V}{E_\l}$ is right continuous whereas
$\l\mapsto\dastoo{V}{E_\l}$ is not. On the other hand, the family
$ \l\mapsto \bigwedge_{\mu>\l}\dastoo{V}{E_\mu}$ \emph{is} right
continuous.} These spectral families lie in $\PV$ and hence, by
the spectral theorem, define self-adjoint operators in $V$. This
leads to the definition of the two daseinisations of an arbitrary
self-adjoint operator:

{\definition Let $\A$ be an arbitrary self-adjoint operator. Then
the \emph{outer} and $\emph{inner}$ daseinisations of $\A$ are
defined at each stage $V$ as:
\begin{eqnarray}
\dastoo{V}{A}&:=&\int_\mathR \l\,
d\big(\delta^i_V(\hat E^A_{\l}) \big),\label{Def:dastooVA}\\
\dastoi{V}{A}&:=&\int_{\mathR}\l\, d
\big(\bigwedge_{\mu>\l}\delta^o_V(\hat E^A_{\mu})\big),
                                            \label{Def:dastoiVA}
\end{eqnarray}
respectively. }

Note that for all $\l\in\mathR$, and for all stages $V$, we have
\begin{equation}
       \dastoi{V}{E_\l}\preceq\bigwedge_{\mu>\l}\dastoo{V}{E_\mu}
                                            \label{dasiE<dasioE}
\end{equation}
and hence, for all $V$,
\begin{equation}
       \dastoi{V}{A}\preceq_s\dastoo{V}{A}.
\end{equation}
This explains why the `$i$' and `$o$' superscripts in
\eqs{Def:dastooVA}{Def:dastoiVA} are defined the way round that
they are.

Both outer daseinisation \eq{Def:dastooVA} and inner daseinisation
\eq{Def:dastoiVA} can be used to `adapt' a self-adjoint operator
$\A$ to contexts $V\in\Ob{\V{}}$ that do not contain $\A$. (On the
other hand, if $\A\in V$, then $\dastoo{V}{A}=\dastoi{V}{A}=\A$.)

\subsubsection{Properties of daseinisation.}
We will now list  some useful properties of daseinisation.

{\bf 1.} It is clear that the outer, and inner, daseinisation
operations can be extended to situations where the self-adjoint
operator $\A$ does not belong to $\BH_\sa$, or where $V$ is not an
{\em abelian} subalgebra of $\BH$. Specifically, let $\mathcal{N}$
be an arbitrary,
 unital von Neumann algebra, and let $\mathcal
S\subset\mathcal{N}$ be a proper unital von Neumann subalgebra.
Then outer and inner daseinisation can be defined as the mappings
\begin{eqnarray}
   \daso:\mathcal{N}_\sa&\map&\mathcal{S}_\sa           \nonumber\\
        \A&\mapsto&\int_\mathR \l\, d
        \big(\dasi_{\mathcal{S}}(\hat E^A_{\l}) \big),\\[8pt]
   \dasi:\mathcal{N}_\sa&\map&\mathcal{S}_\sa   \nonumber\\
         \A&\mapsto&\int_{\mathR}\l\,
d\big(\bigwedge_{\mu>\l}\daso_{\mathcal{S}}(\hat E^A_{\mu})\big).
\end{eqnarray}

A particular case is  $\mathcal{N}=V$ and $\mathcal{S}=V^{\prime}$
for two contexts $V,V^{\prime}$ such that $V^{\prime}\subset V$.
Hence,  a self-adjoint operator can be restricted from one context
to a sub-context.

For the moment, we will let $\mathcal{N}$ be an arbitrary unital
von Neumann algebra, with $\mathcal{S}\subset\mathcal{N}$.

{\bf 2.} By construction,
\begin{equation}
  \dastoo{\mathcal S}{A}=\bigwedge\{\hat B\in
                \mathcal{S}_\sa\mid\hat B\succeq_s\A\},
\end{equation}
where the minimum is taken with respect to the spectral order;
\ie\ $\dastoo{\mathcal S}{A}$ is the smallest self-adjoint
operator in $\mathcal S$ that is spectrally larger than (or equal
to) $\A$. This implies $\dastoo{\mathcal S}{A}\succeq\A$ in the
usual order. Likewise,
\begin{equation}
  \dastoi{\mathcal S}{A}=
  \bigvee\{\hat B\in\mathcal S\mid\hat B\preceq_s\A\},
\end{equation}
so $\dastoi{\mathcal S}{A}$ is the largest self-adjoint operator
in $\mathcal S$ spectrally smaller than (or equal to) $\A$, which
 implies $\dastoi{\mathcal S}{A}\preceq\A$.

{\bf 3.} In general, neither $\dastoo{\mathcal S}{A}$ nor
$\dastoi{\mathcal S}{A}$ can be written as Borel functions of the
operator $\A$, since daseinisation changes the elements of the
spectral family, while a function merely `shuffles them around'.

{\bf 4.} Let $\A\in\mathcal N$ be self-adjoint. The spectrum,
$\sp{A}$, consists of all $\l\in\mathR$ such that the spectral
family $\{\hat E^A_\l\}_{\l\in\mathR}$ is non-constant on any
neighbourhood of $\l$. By definition, outer daseinisation of $\A$
acts on the spectral family of $\A$ by sending $\hat E^A_\l$ to
$\hat E^{\dastoo{\mathcal S}{A}}_\l=\delta^i(\hat
E^A_\l)_{\mathcal S}$. If $\{\hat E^A_\l\}_{\l\in\mathR}$ is
constant on some neighbourhood of $\l$, then the spectral family
$\{\hat E^{\dastoo{\mathcal S}{A}}_\l\}_{\l\in\mathR}$ of
$\dastoo{\mathcal S}{A}$ is also constant on this neighbourhood.
This shows that
\begin{equation}
    {\rm sp}(\dastoo{\mathcal S}{A})\subseteq\sp{A}
\end{equation}
for all self-adjoint operators $\A\in\mathcal{N}_\sa$ and all von
Neumann subalgebras $\mathcal S$. Analogous arguments apply to
inner daseinisation.

Heuristically, this result implies that the spectrum of the
operator $\dastoo{\mathcal S}{A}$ is more degenerate than that of
$\A$; \ie\ the effect of daseinisation is to `collapse'
eigenvalues.

{\bf 5.} Outer and inner daseinisation are both non-linear
mappings. We will show this for projections explicitly. For
example, let $\hat{Q}:=\hat{1}-\P$. Then
$\delta^{o}(\hat{Q}+\P)_{\mathcal S}= \dastoo{\mathcal
S}{1}=\hat{1}$, while $\delta^{o}(\hat{1}-\P)_{\mathcal
S}\succ\hat{1}-\P$ and $\dastoo{\mathcal S}{P}\succ\P$ in general,
so $\delta^{o}(\hat {1}-\P)_{\mathcal S}+\dastoo{\mathcal S}{P}$
is the sum of two non-orthogonal projections in general (and hence
not equal to $\hat{1}$). For inner daseinisation, we have
$\delta^{i}(\hat{1}-\P)_{\mathcal S}\prec\hat{1}-\P$ and
$\dastoi{\mathcal S}{P}\prec\P$ in general, so $\delta^{i}
(\hat{1}-\P)_{\mathcal S}+\dastoi{\mathcal S}{P}\prec\hat{1}
=\delta^{i}(\hat{1}-\P+\P)_{\mathcal S}$ in general.

{\bf 6.} If $a\geq 0$, then $\delta^o(a\A)_{\mathcal
S}=a\dastoo{\mathcal S}{A}$ and $\delta^i(a\A)_{\mathcal
S}=a\dastoi{\mathcal S}{A}$. If $a<0$, then
$\delta^o(a\A)_{\mathcal S}=a\dastoi{\mathcal S}{A}$ and
$\delta^i(a\A)_{\mathcal S}=a\dastoo{\mathcal S}{A}$. This is due
the behaviour of spectral families under the mapping $\A\mapsto
-\A$.

{\bf 7.} Let $\A$ be a self-adjoint operator, and let $\hat
E[A\leq\l]=\hat E^A_\l$ be an element of the spectral family of
$\A$. From \eq{Def:dastooVA} we get
\begin{equation}
\hat E[\delta^o_{\mathcal S}(A)\leq\l]=\delta^i_{\mathcal
S}\big(\hat E[A\leq\l])
\end{equation}
and then
\begin{eqnarray}
     \hat E[\dastoo{{\mathcal S}}{A}>\l] &=& \hat 1-
     \hat E[\dastoo{{\mathcal S}}{A}\leq\l]\\
      &=&\hat 1-\delta^i_{\mathcal S}\big(\hat E[A\leq\l]\big)\\
     &=& \delta^o_{\mathcal S}\big(\hat 1-\hat E[A\leq\l]\big)
                                       \label{Eoi>doE=}
\end{eqnarray}
where we have used the general result that, for any projection
$\P$, we have $\hat 1-\dastoi{{\mathcal S}}{P}=\delta^o_{\mathcal
S}(\hat 1-\P)$. Then, \eq{Eoi>doE=} gives
\begin{equation}
        \hat E[\dastoo{{\mathcal S}}{A}>\l]=
        \delta^o\big(\hat E[A>\l]\big)_{\mathcal S}.
                                \label{E[dA>l]=}
\end{equation}

\subsubsection{The de Groote Presheaves}  We know that
$V\mapsto \dastoo{V}{P}$ and $V\mapsto\dastoi{V}{P}$ are global
elements of the outer presheaf, $\G$, and inner presheaf, $\H$,
respectively. Using the daseinisation operation for self-adjoint
operators, it is straightforward to construct analogous presheaves
for which $V\mapsto\dastoo{V}{A}$ and $V\mapsto\dastoi{V}{A}$ are
global elements. One of these presheaves was  briefly considered
in \cite{deG05}. We call these the `de Groote presheaves' in
recognition of the importance of de Groote's work.

{\definition The {\em outer de Groote presheaf}, $\dG$, is defined
as follows:
\begin{enumerate}
\item[(i)] On objects $V\in\Ob{\V{}}$:  $\dG_V:=V_{\rm sa}$, the
collection of self-adjoint members of $V$.

\item[(ii)] On morphisms $i_{V^{\prime}V}:V^{\prime }\subseteq V:$
The mapping $\dG(i_{V^{\prime}\, V}):\dG_V \map\dG_{V^{\prime}}$
is given by
\begin{eqnarray}
        \dG(i_{V^{\prime}\, V})(\A )&:=&\dastoo{V^{\prime}}{A}\\
&=&\int_\mathR \l\,d\big(\delta^i(\hat E^A_\l)_{V^{\prime}}\big)\\
&=&\int_\mathR \l\,d\big(\H(i_{V^\prime\,V})(\hat E^A_\l)\big)
\end{eqnarray}
for all $\A\in\dG_V$.
\end{enumerate}
} Here we used the  fact that the restriction mapping
$\H(i_{V^\prime\,V})$ of the inner presheaf $\H$ is the inner
daseinisation of projections
$\delta^i:\PV\map\mathcal{P}(V^{\prime})$.

{\definition The {\em inner de Groote presheaf}, $\dH$, is defined
as follows:
\begin{enumerate}
\item[(i)] On objects $V\in\Ob{\V{}}$:  $\dH_V:=V_{\rm sa}$, the
collection of self-adjoint members of $V$.

\item[(ii)] On morphisms $i_{V^{\prime}V}:V^{\prime }\subseteq V:$
The mapping $\dH(i_{V^{\prime}\, V}):\dH_V \map\dH_{V^{\prime}}$
is given by
\begin{eqnarray}
     \dH(i_{V^{\prime}\, V})(\A)&:=&\dastoi{V^{\prime}}{A}\\
     &=&\int_\mathR \l\,  d\big(\bigwedge_{\mu>\l}
     (\delta^o(\hat E^A_\mu)_{V^{\prime}}\big)\\
   &=&\int_\mathR \l\,  d\big(\bigwedge_{\mu>\l}
   (\G(i_{V^\prime\,V})(\hat E^A_\mu)\big)
\end{eqnarray}
for all $\A\in\dG_V$ (where $\G(i_{V^\prime\,V})=\delta^o:
\PV\map\mathcal{P}(V^{\prime})$).
\end{enumerate}
}

It is now clear that, by construction,
$\dastoo{}{A}:=V\mapsto\dastoo{V}{A}$ is a global element of
$\dG$, and $\dastoi{}{A}:=V\mapsto\dastoi{V}{A}$ is a global
element of $\dH$.

De Groote found an example of an element of $\Ga\dG$ that is
\emph{not} of the form $\dastoo{}{A}$ (as mentioned in
\cite{deG05}). The same example can be used to show that there are
global elements of the outer presheaf $\G$ that are not of the
form $\dastoo{}{P}$ for any projection $\P\in\PH$.

\section{The Presheaves $\SpA$ and $\SR$}
\label{Sec:psSR}
\subsection{Background to the Presheaf $\SR$}
Our goal now is to construct a `quantity-value' presheaf $\ps{\R}$
with the property that inner and/or outer daseinisation of an
self-adjoint operator $\A$ can be used to define an arrow, \ie\ a
natural transformation, from $\Sig$ to $\ps{\R}$.\footnote{In
fact, we will define several closely related presheaves that can
serve as a quantity-value object.}

The arrow corresponding to a self-adjoint operator $\A\in\BH$ is
denoted for now by $\breve{A}:\Sig\map\ps{\R}$. At each stage $V$,
we need a mapping
\begin{eqnarray}
              \breve{A}_V:\Sig_V &\map& \ps{\R}_V\\
              \l &\mapsto& \breve{A}_V(\l)
\end{eqnarray}
and we assume that this mapping is given by  evaluation on a
suitable member of $V$. More precisely,  $\l\in\Sig_V$ is a
spectral element\footnote{In this context, by a `spectral
element', $\l\in\Sig_V$ of $V$, we mean a multiplicative, linear
functional $\l:V\map\mathC$ with $\l(\hat 1)=1$.} of $V$ and hence
can be evaluated on operators lying in $V$. And, while $\A$ will
generally not lie in $V$, both the inner daseinisation
$\dastoi{V}{A}$ and the outer daseinisation $\dastoo{V}{A}$ do.

Let us start by considering the operators $\dastoo{V}{A}$, $V\in
\Ob{\V{}}$. Each of these is a self-adjoint operator in the
commutative von Neumann algebra $V$, and hence, by the spectral
theorem, can be represented by a function, (the Gel'fand
transform) $\GT{\dastoo{V}{A}}:\Sig_V\map {\rm
sp}(\dastoo{V}{A})$, with values in the spectrum ${\rm
sp}(\dastoo{V}{A})$ of the self-adjoint operator $\dastoo{V}{A}$.
Since the spectrum of a self-adjoint operator is a subset of
$\mathR$, we can also write $\GT{\dastoo{V}{A}}:\Sig_V\map\mathR$.
The question now is whether the collection of maps
$\GT{\dastoo{V}{A}}:\Sig_V\map\mathR$, $V\in\Ob{\V{}}$, can be
regarded as an arrow from $\Sig$ to some presheaf $\ps{\R}$.

To answer this we need to see how these operators behave as we go
`down a chain' of subalgebras $V^\prime\subseteq V$. The first
remark is that if $V^\prime\subseteq V$ then
$\dastoo{V^\prime}{A}\succeq\dastoo{V}{A}$. When applied to the
Gel'fand transforms, this leads to the equation
\begin{equation}
\GT{\dastoo{V^\prime}{A}}(\l|_{V^{\prime}})\ge
\GT{\dastoo{V}{A}}(\l)
                        \label{dasBVVprime>V}
\end{equation}
for all $\l\in\Sig_V$, where $\l|_{V^\prime}$ denotes the
restriction of the spectral element $\l\in\Sig_V$ to the
subalgebra $V^\prime\subseteq V$. However, the definition of the
spectral presheaf is such that
$\l|_{V^\prime}=\Sig(i_{V^\prime\,V})(\l)$, and hence
\eq{dasBVVprime>V} can be rewritten as
\begin{equation}
\GT{\dastoo{V^\prime}{A}}\big(\Sig(i_{V^\prime\,V})(\l) \big)
        \geq \GT{\dastoo{V}{A}}(\l) \label{dasBVVprime>V2}
\end{equation}
for all $\l\in\Sig_V$.

It is a standard result that the real-number object,
$\ps{\mathR}$, in a presheaf topos $\Set^{\mathcal C^{op}}$ is the
\emph{constant} functor from $\mathcal C$ to $\mathR$ \cite{MM92}.
It follows that the family of Gel'fand transforms,
$\GT{\dastoo{V}{A}}$, $V\in\Ob{\V{}}$,  of the daseinised
operators $\dastoo{V}A$, $V\in\Ob{\V{}}$, cannot define an arrow
from $\Sig$ to $\ps{\mathR}$, as this would require an equality in
\eq{dasBVVprime>V2}, which is not true. Thus the quantity-value
presheaf, $\ps{\R}$, in the topos $\SetH{}$ is \emph{not} the
real-number object $\ps{\mathR}$, although clearly $\ps{\R}$ has
\emph{something} to do with the real numbers. We must take into
account the growth of these real numbers as we go from $V$ to
smaller subalgebras $V^{\prime}$. Similarly, if we consider inner
daseinisation, we get a series of falling real numbers.

\subsection{Definition of the Presheaves $\SpA$ and $\SR$}
\label{SubSec:SR} The inapplicability of the real-number object
$\underline{\mathR}$ may seem strange at first,\footnote{Indeed,
it puzzled us for a while!} but actually it is not that
surprising. Because of the Kochen-Specker theorem, we do not
expect to be able to assign (constant) real numbers as values of
physical quantities, at least not globally. Instead, we draw on
some recent results of M. Jackson \cite{Jac06}, obtained as part
of his extensive study of measure theory on a topos of presheaves.
Here, we use a single construction in Jackson's thesis: the
presheaf of `order-preserving functions' over a partially ordered
set---in our case, $\V{}$. In fact, we will need both
order-reversing and order-preserving functions.

\begin{definition}
Let $(\mathcal{Q},\preceq)$ and $(\mathcal{P},\preceq)$ be
partially ordered
sets. A function%
\begin{equation}
                \mu:\mathcal{Q}\map\mathcal{P}%
\end{equation}
is  \emph{order-preserving} if $q_{1}\preceq q_{2}$ implies $\mu
(q_{1})\preceq\mu(q_{2})$ for all $q_1,q_2\in\mathcal Q$. It is
\emph{order-reversing} if $q_{1}\preceq q_{2}$ implies
$\mu(q_{1})\succeq\mu(q_{2})$.  We denote by
$\mathcal{OP(Q},\mathcal{P)}$ the set of order-preserving
functions $\mu:\mathcal{Q}\map\mathcal{P}$, and by
$\mathcal{OR(Q},\mathcal{P)}$ the set of order-reversing
functions.
\end{definition}
\noindent We note that if $\mu$ is order-preserving, then $-\mu$
is order-reversing, and vice versa.

Adapting Jackson's definitions slightly, if $\cal P$ is any
partially-ordered set, we have the following. {\definition The
\emph{${\cal P}$-valued presheaf, $\ps{{\cal P}}^\succeq$, of
order-reversing functions over $\V{}$} is defined as follows:
\begin{enumerate}
\item[(i)] On objects $V\in\Ob{\V{}}$:
\begin{equation}
     \ps{{\cal P}}^\succeq_V:=\{\mu:\downarrow\!\!V\map {\cal P}\mid
     \mu\in\mathcal{OR}(\downarrow\!\! V,{\cal P}\}
                                                    \label{Def:PGe}
\end{equation}
where $\downarrow\!\!V\subset\Ob{\V{}}$ is the set of all unital
von Neumann subalgebras of $V$.

\item[(ii)] On morphisms $i_{V^{\prime}V}:V^{\prime }\subseteq V:$
The mapping $\ps{{\cal P}}^\succeq(i_{V^{\prime}\, V}):\ps{{\cal
P}}^\succeq_V \map \ps{{\cal P}}^\succeq_{V^{\prime}}$ is given by
\begin{equation}
   \ps{{\cal P}}^\succeq(i_{V^{\prime}\, V})(\mu):=
   \mu_{|_{V^\prime}}
\end{equation}
where $\mu_{|_{V^\prime}}$ denotes the restriction of the function
$\mu$ to $\downarrow\!\!V^\prime\subseteq\downarrow\!\!V$.
\end{enumerate}
} \noindent Jackson uses order-preserving functions with
$\mathcal{P} := [0,\infty)$ (the non-negative reals), with the
usual order $\leq$.

Clearly, there is an analogous definition of the ${\cal P}$-valued
presheaf, $\ps{{\cal P}}^\preceq$, of order-preserving functions
from $\downarrow\!\! V$ to ${\cal P}$. It can be shown that
$\ps{{\cal P}}^\succeq$ and $\ps{{\cal P}}^\preceq$ are isomorphic
objects in $\SetH{}$.

Let us first consider $\ps{{\cal P}}^\succeq$. For us, the key
examples for the partially ordered set ${\cal P}$ are (i)
$\mathR$, the real numbers with the usual order $\leq$, and (ii)
$\rm{sp}(\A)\subset\mathR$, the spectrum of some bounded
self-adjoint operator $\A$, with the order $\leq$ inherited from
$\mathR$. Clearly, the associated presheaf $\SpA$ is a sub-object
of the presheaf $\SR$.

Now let $\A\in\BH_\sa$, and let $V\in \Ob{\V{}}$. Then to each
$\l\in\Sig_{V}$ there is associated the function
\begin{equation}
\dasBVo{V}{A}(\l):\downarrow\!\!V\map \sp{A},
\end{equation}
given by
\begin{eqnarray}
\left(\dasBVo{V}{A}(\l)\right)  (V^{\prime})
    &:=& \GT{\dastoo{V^{\prime}}{A}}(\Sig (i_{V^{\prime}V})(\l))\\
    &=& \GT{\dastoo{V^{\prime}}{A}}(\l|_{V^{\prime}})\\
    &=& \l|_{V^{\prime}}(\dastoo{V^{\prime}}{A})\\
    &=& \l(\dastoo{V^{\prime}}{A})
\end{eqnarray}
for all $V^{\prime}\subseteq V$. We note that as $V^{\prime}$
becomes smaller, $\dastoo{V^{\prime}}{A}$ becomes larger (or stays
the same) in the spectral order, and hence in the usual order on
operators. Therefore, $\dasBVo{V}{A}(\l):\downarrow\!\!
V\map\sp{A}$ is an \emph{order-reversing function}, for each
$\l\in\Sig_V$.

Let
\begin{eqnarray}
                \dasBVo{V}{A}:\Sig_V &\map& \SpA_V\\
                \l &\mapsto& \dasBVo{V}{A}(\l)
\end{eqnarray}
denote the set of order-reversing functions from $\downarrow\!\!
V$ to $\sp{A}$ obtained in this way.

\begin{theorem}\label{Th:ST}
The mappings $\dasBVo{V}{A}$, $V\in\Ob{\V{}}$, are the components
of a natural transformation/arrow $\dasBo{A}:\Sig\map\SpA$.
\end{theorem}

\begin{proof}
We only have to prove that, whenever $V^{\prime}\subset V$, the
diagram
\begin{center}
\setsqparms[1`1`1`1;1000`700]                                 
\square[\Sig_V`\SpA_V`\Sig_{V^{\prime}}`\SpA_{V^{\prime}};    
\dasBVo{V}{A}```\dasBVo{V^{\prime}}{A}]                       
\end{center}

\noindent commutes. Here, the vertical arrows are the restrictions
of the relevant presheaves from the stage $V$ to $V^\prime\subset
V$.

In fact, the commutativity of the diagram follows directly from
the definitions. For each $\l \in\Sig_{V}$, the composition of the
upper arrow and the right vertical arrow gives
\begin{equation}
    (\dasBVo{V}{A}(\l))|_{V^{\prime}}=
    \dasBVo{V^{\prime}}{A}(\l|_{V^{\prime}}),
\end{equation}
which is the same function that we get by first restricting $\l$
from $\Sig_V$ to $\Sig_{V^{\prime}}$ and then applying
$\dasBVo{V^{\prime}}{A}$.
\end{proof}

In this way, to each physical quantity $\A$ in quantum theory
there is assigned a natural transformation $\dasBo{A}$ from the
state object $\Sig$ to the presheaf $\SpA$. Since $\SpA$ is a
sub-object of $\SR$ for each $\A$, $\dasBo{A}$ can also be seen as
a natural transformation/arrow from $\Sig$ to $\SR$. Hence the
presheaf $\SR$ in the topos $\SetH{}$ is one candidate for the
quantity-value object of quantum theory.

In the Appendix, we show that the mapping
\begin{eqnarray}
    \theta:\BH_\sa &\map& \rm {Hom}_{\SetH{}}(\Sig,\SR)\\
                \A &\mapsto& \dasBo{A}
\end{eqnarray}
is injective. Furthermore, if $S$ denotes our quantum system,
then, on the level of the language $\L{S}$, we expect the mapping
$A\map\hat A$ to be injective, where $A$ is a function symbol of
signature $\Si\map\R$. It follows that we have obtained a a
faithful representation of these function symbols by arrows
$\dasBo{A}:\Sig\map\SR$ in the topos $\SetH{}$.

Similarly, there is an order-preserving function
\begin{equation}
                \dasBVi{V}{A}(\l):\downarrow\!\! V\map\sp{A},
\end{equation}
that is defined for all $V^{\prime}\subseteq V$ by
\begin{eqnarray}
   \left(  \dasBVi{V}{A}(\l)\right)  (V^{\prime})
 &=& \GT{\delta^{i}(\A)_{V^{\prime}}}(\Sig(i_{V^{\prime}V})(\l))\\
 &=& \l(\delta^{i}(\A)_{V^{\prime}}).
\end{eqnarray}
Since $\dastoi{V^{\prime}}{A}$ becomes smaller (or stays the same)
as $V^{\prime}$ gets smaller, $\dasBVi{V}{A}(\l)$ indeed is an
\emph{order-preserving} function from $\downarrow\!\! V$ to
$\sp{A}$ for each $\l\in\Sig_V$.

Clearly, we can use these functions to define a natural
transformation $\dasBi{A}:\Sig\map\OP$ from the spectral presheaf,
$\Sig$, to the presheaf $\OP$ of real-valued, order-preserving
functions on $\downarrow\!\! V$. The components of $\dasBi{A}$ are
\begin{eqnarray}
\dasBVi{V}{A}:\Sig_V &\map& \ps{\rm{sp}(\A)^\preceq}_V\\
                \l &\mapsto& \dasBVi{V}{A}(\l).
\end{eqnarray}

The functions obtained from inner and outer daseinisation can be
combined to give yet another  presheaf, and one that will be
particularly useful for the physical interpretation of these
constructions. The general definition is:
\begin{definition}
Let $\mathcal{P}$ be a partially-ordered set. The
\emph{$\mathcal{P}$-valued presheaf, $\PR{\mathcal P}$, of
order-preserving and order-reversing functions on $\V{}$} is
defined as follows:

(i) On objects $V\in Ob(\V{})$:
\begin{equation}
  \ps{\mathcal{P}^{\leftrightarrow}}_{V}
  :=\{(\mu,\nu)\mid\mu\in\mathcal{OP}(\downarrow\!\!V,\mathcal{P}),
\nu\in\mathcal{OR}(\downarrow\!\!V,\mathcal{P})\},
\end{equation}
where $\downarrow\!\! V\subset\Ob{\V{}}$ is the set of all
subalgebras $V^{\prime}$ of $V$.

(ii) On morphisms $i_{V^{\prime}V}:V^{\prime}\subseteq V$:
\begin{eqnarray}
  \ps{\mathcal{P}^{\leftrightarrow}}(i_{V^{\prime}V}):\ps
 {\mathcal{P}^{\leftrightarrow}}_{V}  &\longrightarrow&
 \ps{\mathcal{P}^{\leftrightarrow}}_{V^{\prime}}\\
 (\mu,\nu)  &\longmapsto& (\mu|_{V^{\prime}},\nu|_{V^{\prime}}),
\end{eqnarray}
where $\mu|_{V^{\prime}}$ denotes the restriction of $\mu$ to
$\downarrow\!\! V^{\prime}\subseteq\downarrow\!\! V$, and
analogously for $\nu|_{V^{\prime}}$.
\end{definition}

As we will see shortly, the presheaf, $\PR{\mathR}$, of
order-preserving and order-reversing, real-valued functions is
closely related to the `$k$-extension' of the presheaf $\SR$.

Now let
\begin{equation}
  \dasBV{V}{A}:=\left(  \dasBVi{V}{A}(\cdot),\dasBVo{V}{A}
  (\cdot)\right):
  \Sig_{V}\map\PR{\mathR}_{V}
\end{equation}
denote the set of all pairs of order-preserving and
order-reversing functions from $\downarrow\!\! V$ to $\mathR$ that
can be obtained from inner and outer daseinisation. It is easy to
see that we have the following result:
\begin{theorem}\label{Th:ST_3}
The mappings $\dasBV{V}{A}$, $V\in \Ob{\V{}}$, are the components
of a natural transformation $\dasB{A}:\Sig\map\PR{\mathR}$.
\end{theorem}

\subsection{Inner and Outer Daseinisation from Functions on Filters}
There is a close relationship between inner and outer
daseinisation, and certain functions on the dual ideals/filters in
the projection lattice $\PH$ of $\BH$. We only give a very brief
sketch here: details can be found in de Groote's work
\cite{deG05,deG05b}, the article \cite{Doe05b}, and a forthcoming
paper \cite{Doe07}. This subsection serves as a preparation for
the physical interpretation of the arrows
$\dasB{A}:\Sig\map\PR{\mathR}$.

\paragraph{Spectral elements and ultrafilters.} Let $V\in
\Ob{\V{}}$, and let $\l\in\Sig_{V}$ be a spectral element of the
von Neumann algebra $V$. For all projections
$\P\in\mathcal{P}(V)$, we have
\begin{equation}
    \l(\P)=\l(\P^{2})=\l(\P)\l(\P),
\end{equation}
and so $\l(\P)\in\{0,1\}$. Moreover, $\l(\hat{0})=0$,
$\l(\hat{1})=1$, and if $\l(\P)=0$, then $\l (\hat{1}-\P)=1$
(since $\l(\P)+\l(\hat
{1}-\P)=\l(\hat{1})$). Hence, for each $\P%
\in\mathcal{P}(V)$ we have either $\l(\P)=1$ or $\l
(\hat{1}-\P)=1$. The family
\begin{equation}
    F_{\l}:=\{\P\in\mathcal{P}(V)\mid\l(\P)=1\}
\end{equation}
is an ultrafilter in $\mathcal{P}(V)$.\footnote{Let $\mathbb{L}$
be a lattice with zero element $0$. A subset $F\subset\mathbb{L}$
is a `filter base' if (i) $0\notin F$ and (ii) for all $a,b\in F$,
there is some $c\in F$ such that $c\leq a\wedge b$. A subset
$D\subset\mathbb{L}$ is called a `(proper) dual ideal' or a
`filter' if (i) $0\notin D$, (ii) for all $a,b\in D$, $a\wedge
b\in D$ and (iii) $a\in D$ and $b>a$ implies $b\in D$. A maximal
dual ideal/filter $F$ in a complemented, \emph{distributive}
lattice $\mathbb{L}$ is called an `ultrafilter'. It has the
property that for all $a\in\mathbb{L}$, either $a\in F$ or
$a^{\prime}\in F$, where $a^{\prime}$ is the complement of $a$.}
Conversely, each $\l\in\Sig_{V}$ is uniquely determined by the set
$\{\l(\P)\mid\P\in\mathcal{P}(V)\}$ and hence by an ultrafilter in
$\mathcal{P}(V)$. This shows that there is a bijection between the
set $\mathcal{Q}(V)$ of ultrafilters in $\mathcal{P}(V)$ and the
Gel'fand spectrum $\Sig_{V}$.

Given a filter base $F$ in $\PH$, the \emph{cone over }$F$\emph{
in }$\BH$ is defined as
\begin{equation}
 \mathcal{C}_{\BH}(F):=\{\hat{Q}\in\PH\mid
\exists\P\in F:\P\preceq\hat{Q}\}.
\end{equation}
This is the smallest filter in $\PH$ that contains the filter base
$F$. We write $\mathcal{C}(F)$ for $\mathcal{C}_{\BH}(F)$.

\paragraph{Observable and antonymous functions.} Let $\mathcal{N}$ be
a unital von Neumann algebra, and let $\mathcal{D(N)}$ be the set
of filters in the projection lattice $\mathcal{P(N)}$ of
$\mathcal{N}$. De Groote has shown \cite{deG05b} that to each
self-adjoint operator $\A\in\mathcal{N}$, there corresponds a,
so-called, `observable function' $f_{\A}:\mathcal{D(N)}
\map\sp{A}$. If $\mathcal{N}$ is abelian, $\mathcal{N}=V$, then
$f_{\A}|_{\mathcal{Q}(V)}$ is just the Gel'fand transform of $\A$.
However, it is striking that $f_{\A}$ can be defined even if
$\mathcal{N}$ is non-abelian; for us, the important example is
$\mathcal{N}=\BH$.

If $\{\hat{E}_{\mu}^{A}\}_{\mu\in\mathR}$ is the spectral family
of $\A$, then $f_{\A}$ is defined as
\begin{eqnarray}
        f_{\A}:\mathcal{D(N)} &\map& \sp{A}\nonumber\\
        D &\mapsto& \inf\{\mu\in\mathR\mid\hat{E}_{\mu}^A\in D\}.
\end{eqnarray}
Conversely, given a bounded function $f:\mathcal{D(H)}\map\mathR$
with certain properties, one can find a unique self-adjoint
operator $\A\in\BH$ such that $f=f_{\A}$.

Prop. 3.1 in \cite{deG05} shows that, for all $V\in\Ob{\V{}}$ and
all filters $D$ in $\mathcal{D}(V)$,
\begin{equation}\label{Eq_f_dastooVA(I)=f_A(C(I))}
                f_{\dastoo{V}{A}}(D)=f_{\A}(\mathcal{C(I)}).
\end{equation}
We saw above that to each $\l\in\Sig_{V}$ there corresponds a
unique ultrafilter $F_{\l}\in\mathcal{Q}(V)$.\footnote{De Groote
calls the maximal dual ideals (\ie\ maximal filters) in an
arbitrary lattice the `quasipoints' of the lattice.} Since
$\dastoo{V}{A}\in V$, the observable function $f_{\dastoo{V}{A}}$
is the Gel'fand transform of $\delta ^{o}(\A)_{V}$, and so, on
identifying the ultrafilter $F_{\l}$ with the spectral element
$\l$, we have
\begin{equation}\label{Eq_f_doVA(F_lambda)=lambda(doVA)}
f_{\dastoo{V}{A}}(F_{\l})=\overline{\delta^{o}(\A)_{V}}(\l)
                                =\l(\dastoo{V}{A}).
\end{equation}
From (\ref{Eq_f_dastooVA(I)=f_A(C(I))}) we have
\begin{equation}
   \l(\dastoo{V}{A})=f_{\dastoo{V}{A}}(F_{\l})=
                f_{\A}(\mathcal{C}(F_{\l}))
\end{equation}
for all $V\in\Ob{\V{}}$ and for all $\l\in\Sig_V$. In this sense,
the observable function $f_{\A}$ encodes all the outer
daseinisations $\dastoo{V}{A}$, $V\in\Ob{\V{}}$, of $\A$.

There is also a function, $g_{\A}$, on the filters in $\PH$ that
encodes all the inner daseinisations $\dastoi{V}{A}$, $V\in
\Ob{\V{}}$. This function is given for an arbitrary unital von
Neumann algebra $\mathcal{N}$ by
\begin{eqnarray}
    g_{\A}:\mathcal{D(N)} &\map& \sp{A}\\
    D &\mapsto& \sup\{\l\in\mathR\mid\hat{1}-\hat{E}_{\l}^A\in D\}
\end{eqnarray}
and is called the `antonymous function' of $\A$ \cite{Doe05b}. If
$\mathcal{N}$ is abelian, then $g_{\A}$ is the Gel'fand transform
of $\A$  (and coincides with $f_{\A}$). One can show \cite{Doe07}
that for all $V\in\Ob{\V{}}$ and all filters $D$ in
$\mathcal{D}(V)$,
\begin{equation}\label{Eq_g_dastoiVA(I)=g_A(C(I))}
                g_{\dastoi{V}{A}}(D)=g_{\A}(\mathcal{C(I)}).
\end{equation}
Let $\l\in\Sig_{V}$, and let $F_{\l}\in\mathcal{Q}(V)$ be the
corresponding ultrafilter. Since $\dastoi{V}{A}\in V$, the
antonymous function $g_{\dastoi{V}{A}}$ is the Gel'fand
transform of $\dastoi{V}{A}$, and we have%
\begin{equation}\label{Eq_g_diVA(F_lambda)=lambda(diVA)}
                g_{\dastoi{V}{A}}(F_{\l})=
\overline{\delta^{i}(\A)_{V}}(\l)=\l(\dastoi{V}{A}).
\end{equation}
From (\ref{Eq_g_dastoiVA(I)=g_A(C(I))}), we get
\begin{equation}
 \l(\dastoi{V}{A})=g_{\dastoi{V}{A}}(F_{\l})=
                        g_{\A}(\mathcal{C}(F_{\l}))
\end{equation}
for all $V\in \Ob{\V{}}$ and all  $\l\in\Sig_{V}$. Thus the
antonymous function $g_\A$ encodes all the inner daseinisations
$\dastoi{V}{A}$, $V\in\Ob{\V{}}$, of $\A$.

\subsection{A Physical Interpretation of the Arrow
$\dasB{A}:\Sig\map\PR{\mathR}$}

Let $\ket\psi\in\mathcal{H}$ be a unit vector in the Hilbert space
of the quantum system. The expectation value of a self-adjoint
operator $\A\in \BH$ in the state $\ket\psi$ is given by
\begin{equation}
  \bra\psi\A\ket\psi=
\int_{-||\A||}^{||\A||}\l\, d\bra\psi\hat{E}_{\l}^A\ket\psi.
\end{equation}
There is a maximal filter $F_{\ket\psi}$ in $\PH$,\footnote{Since
$\PH$ is not distributive, $F_{\ket\psi}$ is not an ultrafilter;
\ie\ there are projections $\P\in\PH$ such that neither $\P\in
F_{\ket\psi}$ nor $\hat 1-\P\in F_{\ket\psi}$.} given by
\begin{equation}
F_{\ket\psi}:=\{\P\in\PH\mid\P\succeq\P_{\ket\psi}\},
\end{equation}
where $\P_{\ket\psi}$ is the projection onto the one-dimensional
subspace of $\mathcal H$ generated by $\ket\psi$. As shown in
\cite{Doe05b}, the expectation value $\bra\psi\A\ket\psi$ can be
written as
\begin{equation}
         \bra\psi\A\ket\psi=
       \int_{g_{\A}(F_{\ket\psi})}^{f_{\A}(F_{\ket\psi})}
\l\, d\bra\psi\hat{E}_{\l}^A\ket\psi.
\end{equation}

In an instrumentalist interpretation,\footnote{Which we avoid in
general, of course!} one would interpret $g_{\A}(F_{\ket\psi})$,
resp.\ $f_{\A}(F_{\ket\psi})$, as the smallest, resp.\ largest,
possible result of a measurement of the physical quantity $A$ when
the state is $\ket\psi$. If $\ket\psi$ is an eigenstate of $\A$,
then $\bra\psi\A\ket\psi$ is an eigenvalue of $\A$, and in this
case, $\bra\psi\A\ket\psi\in\sp{A};$  moreover,
\begin{equation}
\bra\psi\A\ket\psi =g_{\A}(F_{\ket\psi})=f_{\A}(F_{\ket\psi}).
\end{equation}
If $\ket\psi$ is not an eigenstate of $\A$, then%
\begin{equation}
g_{\A}(F_{\ket\psi})<\bra\psi\A\ket\psi<f_{\A}(F_{\ket\psi}).
\end{equation}

Let $V$ be an abelian subalgebra of $\BH$ such that $\Sig_V$
contains the spectral element, $\l^{\ket\psi}$, associated with
$\ket\psi$.\footnote{This is the element defined by
$\l^{\ket\psi}(\A):=\bra\psi\A\ket\psi$ for all $\A\in V$.} The
corresponding ultrafilter $\tilde F_{\ket\psi}$ in
$\mathcal{P}(V)$ consists of those projections
$\hat{Q}\in\mathcal{P}(V)$ such that $\hat{Q}\succeq\hat
P_{\ket\psi}$. Hence the cone $\mathcal{C}(\tilde F_{\ket\psi})$
consists of all projections $\hat{R}\in\PH$ such that
$\hat{R}\succeq\hat P_{\ket\psi}$; and so
\begin{equation}
                \mathcal{C}(\tilde F_{\ket\psi})=F_{\ket\psi}.
\end{equation}
This allows us to write the expectation value as
\begin{eqnarray}
   \bra\psi\A\ket\psi
   &=&\int_{g_{\A}(\mathcal{C}
   (\tilde F_{\ket\psi}))}^{f_{\A}(\mathcal{C}
   (\tilde F_{\ket\psi}))}\l\, d\bra\psi\hat{E}_{\l}^A\ket\psi\\
  &=& \int_{g_{\dastoi{V}{A}}
  (\tilde F_{\ket\psi})}^{f_{\dastoo{V}{A}}
  (\tilde F_{\ket\psi})}\l\, d\bra\psi\hat{E}_{\l}^A\ket\psi.
\end{eqnarray}

Equations (\ref{Eq_f_doVA(F_lambda)=lambda(doVA)}) and
(\ref{Eq_g_diVA(F_lambda)=lambda(diVA)}) show that
$f_{\dastoo{V}{A}}(\tilde
F_{\ket\psi})=\bra\psi\dastoo{V}{A}\ket\psi$ and\newline
\mbox{$g_{\delta^{i}(\A)_{V}}(\tilde
F_{\ket\psi})=\bra\psi\dastoi{V}{A}\ket\psi$.} In the language of
instrumentalism, for stages $V$ for which
$\l^{\ket\psi}\in\Sig_{V}$,  the value
$\bra\psi\dastoi{V}{A}\ket\psi\in\sp{A}$ is the smallest possible
measurement result for $\A$ in the quantum state $\ket\psi$; and
$\bra\psi\dastoo{V}{A}\ket\psi\in\sp{A}$ is the largest possible
result.

If $\l\in\Sig_{V}$ is not of the form $\l=\l^{\ket\psi}$, then the
cone $\mathcal{C}(F_{\l})$ over the ultrafilter $F_{\l}$
corresponding to $\l$ cannot be identified with a vector in $\Hi$.
Nevertheless, the quantity $\mathcal{C}(F_{\l})$ is well-defined,
and (\ref{Eq_f_dastooVA(I)=f_A(C(I))}) and
(\ref{Eq_g_dastoiVA(I)=g_A(C(I))}) hold. If we go from $V$ to a
subalgebra $V^{\prime}\subset V$, then $\delta^{i}(\hat
{A})_{V^{\prime}}\preceq\dastoi{V}{A}$ and $\delta^{o}(\hat
{A})_{V^{\prime}}\succeq\dastoo{V}{A}$, hence
\begin{eqnarray}
     \l(\delta^{i}(\A)_{V^{\prime}}) &\leq& \l(\delta^{i}(\A)_{V}),\\
     \l(\delta^{o}(\A)_{V^{\prime}}) &\geq& \l(\delta^{o}(\A)_{V})
\end{eqnarray}
for all $\l\in\Sig_V$.

We can interpret the function
\begin{eqnarray}
  \dasBV{V}{A}:\Sig_{V} &\map& \PR{\mathR}_{V}\\
  \l &\mapsto& \dasBV{V}{A}(\l)=
           \left( \dasBVi{V}{A}(\l),\dasBVo{V}{A}(\l) \right)
\end{eqnarray}
as giving the `spread' or `range' of the physical quantity $A$ at
stages $V^{\prime}\subseteq V$. Each element $\l\in\Sig_{V}$ gives
its own `spread' $\dasBV{V}{A}(\l):\downarrow\!\! V\map
\sp{A}\times\sp{A}$. The intuitive idea is that at stage $V$,
given a point $\l\in\Sig_{V}$, the physical quantity $A$
`spreads over' the subset%
\begin{equation}
  \lbrack\dasBVi{V}{A}(\l)(V),\dasBVo{V}{A}(\l)(V)]\cap\sp{A}=
  \lbrack\l(\dastoi{V}{A}),\l(\delta^{o}(\A))_{V})]\cap\sp{A},
\end{equation}
and for $V^{\prime}\subset V$, over the (potentially larger)
subset
\begin{equation}
  \lbrack\dasBVi{V}{A}(\l)(V^{\prime}),
                        \dasBVo{V}{A}(\l)(V^{\prime})]\cap\sp{A}=
  \lbrack\l(\delta^{i}(\A)_{V^{\prime}},
                     \l(\delta^{o}(\A))_{V^{\prime}}]\cap\sp{A}.
\end{equation}

All this is local in the sense that these expressions are defined
at a stage $V$ and for subalgebras, $V^\prime$, of $V$, where
$\l\in\Sig_{V}$. No similar global construction or interpretation
is possible, since the spectral presheaf $\Sig$ has no global
elements, i.e., no points (while the \emph{set} $\Sig_{V}$ does
have points).

As we go down to smaller subalgebras $V^{\prime}\subseteq V$, the
spread gets larger. This comes from the fact that $\A$ has to be
adopted more and more as we go to smaller subalgebras
$V^{\prime}$. More precisely, $\A$ is approximated from below by
$\delta^{i}(\hat {A})_{V^{\prime}}\in V^{\prime}$ and from above
by $\delta^{o}(\hat {A})_{V^{\prime}}\in V^{\prime}$. This
approximation gets coarser as $V^{\prime}$ gets smaller, which
basically means that $V^{\prime}$ contains less and less
projections.

It should be remarked that $\dasB{A}$ does not assign actual
values to the physical quantity $A$, but rather the possible
\emph{range} of such values; and these are independent of any
state $\ket\psi$.  This is analogous to the classical case where
physical quantities are represented by  real-valued functions on
state space. The range of possible values is state-independent,
but the actual value possessed by  a physical quantity \emph{does}
depend on the state of the system.

\subsection{Properties of $\SR$.}
From the perspective of our overall programme, Theorem \ref{Th:ST}
is a key result and suggests strongly that $\SR$ is an appropriate
choice for the quantity-value object for quantum theory. To
explore this further, we note the following elementary properties
of $\SR$.\footnote{Analogous arguments apply to the presheaves
$\OP$ and $\PR{\mathR}$.}
\begin{enumerate}
\item The presheaf $\SR$ has global elements:  namely,
order-reversing functions on the partially-ordered set $\Ob{\V{}}$
of objects in the category $\V{}$; \ie\ functions
$\mu:\Ob{\V{}}\map\mathR$ such that:
\begin{equation}
   \mbox{For all $V_1,V_2\in\Ob{\V{}}$, }
  V_2\subseteq V_1 \mbox{ implies } \mu(V_2)\geq\mu(V_1).
                        \label{Def:GammaRD}
\end{equation}

\item
\begin{enumerate}
  \item Elements of $\Ga\SR$ can be added: \ie\ if $\mu,\nu\in\Ga\SR$,
  define $\mu+\nu$ at each stage $V$ by
      \begin{equation}
          (\mu+\nu)(V^\prime):=\mu(V^\prime)+\nu(V^\prime)
          \label{Def:mu+nu}
      \end{equation} for all $V^\prime\subseteq V$. Note that if
$V_2\subseteq V_1\subseteq V$, then $\mu(V_1)\leq \mu(V_2)$ and
$\nu(V_1)\leq\nu(V_2)$, and so $\mu(V_1)+\nu(V_1)\leq\mu(V_2)
+\nu(V_2)$. Thus the definition of $\mu+\nu$ in \eq{Def:mu+nu}
makes sense. Obviously, addition is commutative  and associative.

\item  However, it is \emph{not} possible to define `$\mu-\nu$'
       in this way since the difference between two order-reversing
       functions may not be order-reversing. This problem is addressed
       in Section \ref{Sec:PskSR}.

\item A `zero/unit' element can be defined for the additive
structure on $\Ga\SR$ as $0(V):=0$ for all $V\in\Ob{\V{}}$, where
$0$ denotes the function that is constantly $0$ on $\Ob{\V{}}$.

It follows from (a) and (c) that $\Ga\SR$ is a commutative monoid
(\ie\ a semi-group with a unit).
\end{enumerate}

    The commutative monoid structure for $\Ga\SR$ is a
  reflection of the stronger fact that $\SR$ is  a
  \emph{commutative-monoid} object in the topos $\SetH{}$.
  Specifically, there is an arrow $+:\SR\times\SR\map\SR$ defined by
   $+_V(\mu,\nu):=\mu+\nu$ for all $\mu,\nu\in\SR_V$,
   and for all stages $V\in\Ob{\V{}}$. Here, $\mu+\nu$ denotes the
   real-valued function on $\downarrow\!\!V$ defined by
   \begin{equation}
    (\mu+\nu)(V^\prime):=\mu(V^\prime)+\nu(V^\prime)
            \label{Def:mu+nuFn}
    \end{equation}
    for all $V^\prime\subseteq V$.

    \item The real numbers, $\mathR$, form a ring, and so it is
    natural to see if a multiplicative structure can be put on
    $\Ga\SR$. The obvious `definition' would be, for all $V$,
    \begin{equation}
    (\mu\nu)(V):=\mu(V)\nu(V)     \label{Def:ab:=}
    \end{equation}
    for $\mu,\nu\in\Ga\SR$. However, this fails because the right hand
    side of \eq{Def:ab:=} may not be order-reversing. This problem
    arises if $\mu(V)$ and $\nu(V)$ become negative: then, as $V$ gets
    smaller, these numbers get closer and closer to $0$, and then their
    product is a function that is \emph{order-preserving}.
 \end{enumerate}

\subsection{The Representation of Propositions From Pullbacks}
\label{_Sec_RepOfPropositions} In paper I \cite{DI(1)}, we
introduced a simple propositional language, $\PL{S}$, for each
system $S$, and discussed its representations for the case of
classical physics. Then, in \cite{DI(2)} we analysed the, far more
complicated, quantum-theoretical representation of this language
in the set of clopen subsets of the spectral presheaf, $\Sig$, in
the topos $\SetH{}$. This gives a representation of the primitive
propositions $\SAin\De$ as
\begin{equation}
        \piqt(\Ain\De):=\delta^o\big(\hat E[A\in\De]\big)
                                                \label{Def:piAinD}
\end{equation}
where `$\delta^o$' is the (outer) daseinisation operation, and
$\hat E[A\in\De]$ is the spectral projection on the subset
$\De\cap\sp{A}$ of the spectrum, $\sp{A}$, of the self-adjoint
operator $\A$.

We now want to remark briefly on the nature, and representation,
of propositions using the `local' language $\L{S}$.

In any classical representation, $\s$, of $\L{S}$ in $\Set$, the
representation, $\R_\s$, of the quantity-value symbol $\R$ is
always just the real numbers $\mathR$. Therefore, it is simple to
take a subset $\De\subseteq\mathR$ of $\mathR,$ and construct the
propositions $\SAin\De$. In fact, if $A_\s:\Si_\s\map\mathR$ is
the representation of the function symbol $A$ with signature
$\Si\map\R$, then $A_\s^{-1}(\De)$ is a subset of the symplectic
manifold $\Si_\s$ (the representation of the ground type $\Si$).
This subset, $A_\s^{-1}(\De)\subseteq\Si_\s$, represents the
proposition $\SAin\De$ in the Boolean algebra of all (Borel)
subsets of $\Si_\s$.

We should consider the analogue of these steps in the
representation, $\phi$, of the same language, $\L{S}$, in the
topos $\tau_\phi:=\SetH{}$.  In fact, the  issues to be discussed
apply to a representation in \emph{any} topos.

We first note that  if $\Xi $ is a sub-object of $\R_\phi$, and if
$A_\phi:\Si_\phi\map{\mathcal R}_\phi$, then there is an
associated sub-object of $\Si_\phi$, denoted $A_\phi^{-1}(\Xi)$.
Specifically, if $\chi_{\Xi}:\R_\phi\map\O_{\tau_\phi}$ is the
characteristic arrow of the sub-object $\Xi$, then
$A_\phi^{-1}(\Xi)$ is defined to be the sub-object of $\Si_\phi$
whose characteristic arrow is $\chi_{\Xi}\circ
A_\phi:\Si_\phi\map\O_{\tau_\phi}$. These sub-objects are
analogues of the subsets, $A_\s^{-1}(\De)$, of the classical state
space $\Si_\s$: as such, they can represent propositions. In this
spirit, we could denote by $\SAin\Xi$ the proposition which the
sub-object $A_\phi^{-1}(\Xi)$ represents, although, of course, it
would be a mistake to interpret $\SAin\Xi$ as asserting that the
value of something lies in something else: in a general topos,
there are no such values.

In the case of quantum theory, the arrows
$A_\phi:\Si_\phi\map\R_\phi$ are of the form
$\dasBo{A}:\Sig\map\SR$ where $\R_\phi:=\SR$.\footnote{We can also
use the other presheaves defined in Subsection \ref{SubSec:SR},
$\OP$ and $\PR{\mathR}$, as quantity-value objects and pull back
their sub-objects to sub-objects of $\Sig$.} It follows that the
propositions in our $\L{S}$-theory are represented by the
sub-objects $\dasBo{A}^{-1}(\ps{\Xi})$ of $\Sig$, where $\ps{\Xi}$
is a sub-object of $\SR$.

To interpret such propositions, note first that in the
$\PL{S}$-propositions $\SAin\De$,  the range `$\De$' belongs to
the world that is external to the language. Consequently, the
\emph{meaning} of $\De$ is given  independently of $\PL{S}$. This
`externally interpreted' $\De$ is then inserted into the quantum
representation of $\PL{S}$ via the daseinisation of propositions
discussed in paper II.

However, the situation is very different for the
$\L{S}$-propositions $\SAin\Xi$. Here, the quantity `$\Xi$'
belongs to the particular topos $\tau_\phi$, and hence it is
representation dependent. The implication is that the `meaning' of
$\SAin\Xi$  can only be discussed from `within the topos' using
the internal language that is associated with $\tau_\phi$, which,
we recall, carries the translation of $\L{S}$ given by the
topos-representation $\phi$.

From  a conceptual perspective, this situation is  `relational',
with the meanings of the various propositions being determined by
their relations to each other as formulated in the internal
language of the topos. Concomitantly, the  meaning of `truth'
cannot be understood using the \emph{correspondence theory} (much
favoured by instrumentalists) for there is nothing external to
which a proposition can `correspond'. Instead, what is needed is
more like a \emph{coherence} theory of truth in which  a whole
body of propositions is considered together \cite{Gray90}. This is
a fascinating subject, but further discussion  must be deferred to
later work.

\section{The Presheaf $\kSR$}
\label{Sec:PskSR}
\subsection{Some Background Information}
\subsubsection{Preliminary Remarks}
We have shown how each self-adjoint operator, $\A$, on the Hilbert
space $\Hi$ gives rise to an arrow $\dasBo{A}:\Sig\map\SR$ in the
topos $\SetH{}$. Thus, in the topos representation, $\phi$, of
$\L{S}$ for the theory-type `quantum theory', the arrow
$\dasBo{A}:\Sig\map\SR$  is one possible choice\footnote{Another
choice is to use the presheaf $\PR\mathR$ as the quantity-value
object.} for  the representation, $A_\phi:\Si_\phi\map{\mathcal
R}_\phi$, of the function symbol, $A:\Si\map\mathcal R$.

This implies that the quantity-value object, $\R_\phi$, is the
presheaf, $\SR$. However, although such an identification is
 possible, it does impose certain restrictions on the
formalism. These stem from the fact that $\SR$ is only a
\emph{monoid}-object in $\SetH{}$, and $\Ga\SR$ is only a monoid,
whereas the real numbers of standard physics are an abelian group;
indeed, they are a commutative ring.

In standard classical physics, $\Hom{\Set}{\Si_{\s}}{\mathR}$ is
the set of real-valued functions on the manifold $\Si_{\s}$; as
such, it possesses the structure of a commutative ring. On the
other hand, the set of arrows $\Hom{\SetH{}}{\Sig}{\SR}$ has only
the structure of an additive monoid. This additive structure is
defined locally in the obvious way: \ie\ if
$\alpha,\beta\in\Hom{\SetH{}}{\Sig}{\SR}$ we define, for all
stages $V\in\Ob{\V{}}$, and all $\l\in\Sig_V$, (c.f.\
\eq{Def:mu+nuFn})
\begin{equation}
 (\alpha+\beta)_V(\l):=\alpha_V(\l)+\beta_V(\l)
                                        \label{Def:a+b}
\end{equation}
where both sides of \eq{Def:a+b} are elements of $\SR_V$, \ie\
order-reversing functions from $\downarrow\!\!V$ to $\mathR$. It
is clear that the right hand side of \eq{Def:a+b} \emph{is} an
order-reversing function, so that $\alpha+\beta$ is well
defined.\footnote{To avoid confusion we should emphasise that, in
general, the sum $\dasoB{A}+\dasoB{B}$ is \emph{not} equal to
$\daso(\A+\hat B)$.}

Arguably, the fact that $\Hom{\SetH{}}{\Sig}{\SR}$ is only a
monoid\footnote{An internal version of this result would show that
the exponential object $\SR^{\,\Sig}$ is a monoid object in the
topos $\SetH{}$. This could well be true, but we have not studied
it in detail.} is  a weakness in so far as we are trying to make
quantum theory `look' as much like classical physics as possible.
Of course, in more obscure applications such as Planck-length
quantum gravity, the nature of the quantity-value object is very
much open for debate. But when applied to regular physics, we
might like our formalism to look more like classical physics than
the monoid-only structure of $\Hom{\SetH{}}{\Sig}{\SR}$.

There are also more practical reasons for wanting to extend our
current formalism. For example, given the arrows $\dasBo{A^2}$ and
$\dasBo{A}$,  it would be interesting to  define an `intrinsic
dispersion'\footnote{The notation used here is potentially a
little misleading. We have not given any meaning to `$A^2$' in the
language $\L{S}$; \ie\ in its current form, the language does not
give meaning to the square of a function symbol. Therefore, when
we write $\dasBo{A^2}$ this must be understood as being the
Gel'fand transform of the outer daseinisation of the operator
$\A^2$.}:
\begin{equation}
        \nabla(\A):=\dasBo{A^2}-\dasBo{A}^2   \label{Def:nabla}
\end{equation}
but this is meaningless because (i) the square, $\dasBo{A}^2$, of
the arrow $\dasBo{A}:\Sig\map\SR$ is not well-defined; and (ii)
even if it was, the difference between two arrows from $\Sig$ to
$\SR$ is not well-defined. Thus we need  to be able to take the
square of an element in $\Hom{\SetH{}}{\Sig}{\SR}$, and to
subtract such arrows.

The need for a subtraction, i.e. some sort of abelian group
structure on $\SR$, brings to mind the well-known Grothendieck
$k$-construction that is much used in algebraic topology and other
branches of pure mathematics. This gives a way of `extending' an
abelian semi-group to become an abelian group, and we want to see
if this technique can be adapted to the present situation. The
goal is to construct a `Grothendieck completion', $\kSR$, of $\SR$
that is an abelian-group object in the topos $\SetH{}$.

Of course, what we want is more, namely a well-defined operation
of taking squares. We will see that, in  a limited sense, this is
obtained at no extra cost.\footnote{Ideally, we might like $\kSR$
to be a commutative-ring object, but this is not true.}

\subsubsection{The Grothendieck $k$-Construction for a Semi-Group}
Let us briefly review the  Grothendieck construction for  an
abelian monoid $M$.

{\definition A \emph{group completion} of $M$ is an abelian group
$k(M)$ together with a monoid map $\theta:M\map k(M)$ that is
universal. Namely, given any monoid morphism $\phi:M\map G$, where
$G$ is an abelian group, there exists a unique \emph{group}
morphism $\phi^\prime: k(M)\map G$  such that $\phi$ factors
through $\phi^\prime$; \ie\ we have the commutative diagram
\begin{center}
   \Vtrianglep<1`1`-1;400>[M`G`k(M);\phi`\theta`\phi^{\prime}]
\end{center}
with $\phi=\phi^\prime\circ\theta$. }

\noindent It is easy to see that any such $k(M)$ is unique up to
isomorphism.

To prove existence, first take the set of all pairs $(a,b)\in
M\times M$, each of which is to be thought of heuristically as
$a-b$. Then,  note that \emph{if} inverses existed in $M$, we
would have  $a-b=c-d$ if and only if $a+d=c+b$. This suggests
defining an equivalence relation on $M\times M$ in the following
way:
\begin{equation}
  (a,b)\equiv (c,d)  \makebox{ iff $\exists g\in M$ such that }
        a+d+g=b+c+g.     \label{Def:k(M)}
\end{equation}

{\definition The \emph{Grothendieck completion} of a monoid $M$ is
the pair $(k(M),\theta)$ defined as follows:
\begin{enumerate}
\item[(i)] $k(M)$  is the set of equivalence classes $[a,b]$, where
the equivalence relation is defined in \eq{Def:k(M)}. A group law
on $k(M)$ is defined by
\begin{eqnarray}
 &&{\rm (i)}\   [a,b]+[c,d]:=[a+c,b+d],           \\[2pt]
 &&{\rm (ii)}\          0_{k(M)}:=[0_M,0_M],      \\[2pt]
 &&{\rm (iii)}\  -[a,b]:=[b,a],
\end{eqnarray}
where $0_M$ is the unit in the  abelian monoid $M$.

\item[(ii)] The map $\theta:M\map k(M)$ is defined by
\begin{equation}
\theta(a):=[a,0]
\end{equation}
for all $a\in M$.
\end{enumerate}
} It is straightforward to show that (i) these definitions are
independent of the representative elements in the equivalence
classes; (ii) the axioms for a group are satisfied; and (iii) the
map $\theta$ is universal in the sense mentioned above.

It is also clear that $k$ is a \emph{functor} from the category of
abelian monoids to the category of abelian groups. For, if
$f:M_1\map M_2$ is a morphism between abelian monoids,  define
$k(f):k(M_1)\map k(M_2)$ by $k(f)[a,b]:=[f(a),f(b)]$ for all
$a,b\in M_1$.

\subsection{Functions of Bounded Variation and $\Ga\SR$}
These techniques will now be applied to the set, $\Ga\SR$, of
global elements of $\SR$. As was discussed in Section
\ref{SubSec:SR}, global elements of $\SR$ are in one-to-one
correspondence with order-reversing functions on  the category
$\V{}$; \ie\ with functions $\mu:\Ob{\V{}}\map\mathR$ such that,
for all $V_1,V_2\in\Ob{\V{}}$, if $V_2\subseteq V_1$ then
$\mu(V_2)\geq\mu(V_1)$; see \eq{Def:GammaRD}. The monoid law is
given by \eq{Def:mu+nuFn}.

Since $\Ga\SR$ is an abelian monoid, the Grothendieck construction
can be applied to give an abelian group $k(\Ga\SR)$. This is
defined to be the set of equivalence classes $[\l,\ka]$ where
$\l,\ka\in\Ga\SR$, and where $(\l_1,\ka_1)\equiv(\l_2,\ka_2)$ if,
and only if, there exists $\alpha\in\Ga\SR$, such that
\begin{equation}
        \l_1+\ka_2+\alpha=\ka_1+\l_2+\alpha      \label{Def:equivkGR}
\end{equation}

Intuitively, we can think of $[\l,\ka]$ as being `$\l-\ka$', and
embed $\Ga\SR$ in $k(\Ga\SR)$ by $\l\mapsto[\l,0]$. However,
$\l,\ka$ are $\mathR$-valued functions on $\Ob{\V{}}$ and hence,
in this case, the expression `$\l-\ka$' also has a \emph{literal}
meaning: \ie\ as the function $(\l-\ka)(V):=\l(V)-\ka(V)$ for all
$V\in\Ob{\V{}}$.

This is not just a coincidence of notation. Indeed, let
$F\big(\Ob{\V{}},\mathR\big)$ denote the set of all real-valued
functions on $\Ob{\V{}}$. Then we can construct the map,
\begin{eqnarray}
          j:k(\Ga\SR)&\map   & F\big(\Ob{\V{}},\mathR\big)
                                        \label{Def:jk->F} \\
           {[}\l,\ka]&\mapsto&  \l-\ka  \nonumber
\end{eqnarray}
which is  well-defined on equivalence classes.

It is  easy to see that the map in \eq{Def:jk->F} is injective.
This raises the question of  the \emph{image} in
$F\big(\Ob{\V{}},\mathR\big)$ of the map $j$: \ie\ what types of
real-valued function on $\Ob{\V{}}$ can be written as the
difference between two order-reversing functions?

For functions $f:\mathR\map\mathR$, it is a standard result that a
function can be written as the difference between two monotonic
functions if, and only if, it has bounded variation. The natural
conjecture is that a similar result applies here. To show this, we
proceed as follows.

Let $f:\Ob{\V{}}\map\mathR$ be a real-valued function on the set
of objects in the category $\V{}$. At each $V\in\Ob{\V{}}$,
consider a finite chain
\begin{equation}
    C:=\{V_0,V_1,V_2,\ldots,\ V_{n-1},V\mid V_0\subset
    V_1\subset V_2\subset\cdots\subset V_{n-1}\subset V\}
\end{equation}
of proper subsets, and define the \emph{variation} of $f$ on this
chain to be
\begin{equation}
        V_f(C):=\sum_{j=1}^n|f(V_j)-f(V_{j-1})|
\end{equation}
where we set $V_n:=V$. Now take the supremum of $V_f(C)$ for all
such chains $C$. If this is finite, we say that $f$ has a
\emph{bounded variation} and define
\begin{equation}
        I_f(V):=\sup_C V_f(C)
\end{equation}

Then it is clear that (i) $V\mapsto I_f(V)$ is an order-preserving
function on $\Ob{\V{}}$; (ii) $f-I_f$ is an order-reversing
function on $\Ob{\V{}}$; and (iii) $-I_f$ is an order-reversing
function on $\Ob{\V{}}$. Thus, any function, $f$, of bounded
variation can be written as
\begin{equation}
        f\equiv(f-I_f)-(-I_f)
\end{equation}
which is the difference of two order-reversing functions; \ie\ $f$
can be expressed as the difference of two elements of $\Ga\SR$.

Conversely, it is a straightforward modification of the proof for
functions \mbox{$f:\mathR\map\mathR$}, to show that if
$f:\Ob{\V{}}\map\mathR$ is the difference of two order-reversing
functions, then $f$ is of bounded variation. The conclusion is
that $k(\Ga\SR)$ is in bijective correspondence with the set,
$\BV$, of functions  $f:\Ob{\V{}}\map\mathR$ of bounded variation.

\subsection{Taking Squares in $k(\Ga\SR$).}
We can now think of $k(\Ga\SR)$ in two ways: (i) as the set of
equivalence classes $[\l,\ka]$, of elements $\l,\ka\in\Ga\SR$; and
(ii) as the set, $\BV$, of differences $\l-\ka$ of such elements.

As expected, $\BV$ is an abelian group. Indeed: suppose $\alpha
=\l_1-\ka_1$ and $\beta=\l_2-\ka_2$ with $\l_1,\l_2,\ka_1,\ka_2\in
\Ga\SR$, then
\begin{equation}
        \alpha+\beta=(\l_1+\l_2)-(\ka_1+\ka_2)
\end{equation}
Hence $\alpha+\beta$ belongs to $\BV$ since $\l_1+\l_2$ and $\ka_1
+\ka_2$ belong to $\Ga\SR$.

\paragraph{The definition of $[\l,0]^2$.} We will now show how to
take the square of elements of $k(\Ga\SR)$ that are of the form
$[\l,0]$. Clearly, $\l^2$ is well-defined as a function on
$\Ob{\V{}}$, but it may not belong to $\Ga\SR$. Indeed, if
$\l(V)<0$ for any $V$, then the function $V\mapsto\l^2(V)$ can get
smaller as $V$ gets smaller, so it is order-preserving instead of
order-reversing.

This suggests the following strategy. First, define functions
$\l_+$ and $\l_-$ by
\begin{equation}
\l_+(V):= \left\{\begin{array}{ll}
            \l(V) & \mbox{\ if\ $\l(V)\geq 0$} \\[2pt]
            0 & \mbox{\ if\ $\l(V)< 0$}
         \end{array}
        \right.
\end{equation}
and
\begin{equation}
\l_-(V):= \left\{\begin{array}{ll}
            0 & \mbox{\ if\ $\l(V)\geq0$} \\[2pt]
            \l(V) & \mbox{\ if\ $\l(V)<0$.}
         \end{array}
        \right.
\end{equation}
Clearly, $\l(V)=\l_+(V)+\l_-(V)$ for all $V\in\Ob{\V{}}$. Also,
for all $V$, $\l_+(V)\l_-(V)=0$, and hence
\begin{equation}
        \l(V)^2=\l_+(V)^2+\l_-(V)^2 \label{l2=lL2+lR2}
\end{equation}
However, (i) the function $V\mapsto\l_+(V)^2$ is order-reversing;
and (ii) the function $V\mapsto\l_-(V)^2$ is order-preserving. But
then $V\mapsto -\l_-(V)^2$ is order-reversing. Hence, by rewriting
\eq{l2=lL2+lR2} as
\begin{equation}
        \l(V)^2=\l_+(V)^2-(-\l_-(V)^2)
\end{equation}
we see that the function $V\mapsto \l^2(V):=\l(V)^2$ is an element
of $\BV$.

In terms of $k(\Ga\SR)$, we can define
\begin{equation}
        [\l,0]^2:=[\l_+^2, -\l_-^2]
\end{equation}
which belongs to $k(\Ga\SR)$. Hence, although there exist
$\l\in\Ga\SR$ that have no square in $\Ga\SR$, such global
elements of $\SR$ \emph{do} have squares in the $k$-completion,
$k(\Ga\SR)$. On the level of functions of bounded variation, we
have shown that the square of a monotonic (order-reversing)
function is a function of bounded variation.

On the other hand, we cannot take the square of an arbitrary
element $[\l,\ka]\in\Ga\SR$, since the square of a function of
bounded variation need not be a function of bounded
variation.\footnote{We have to consider functions like $(\l_+
+\l_- -(\ka_+ +\ka_-))^2$, which contains terms of the form
$\l_+\ka_-$ and $\l_-\ka_+$: in general, these are neither
order-preserving nor order-reversing.}

\subsection{The Object $\kSR$ in the Topos $\SetH{}$}
\subsubsection{The  Definition of $\kSR$}
The next step is to translate these results about the set
$k(\Ga\SR)$ into the construction of an object $\kSR$ in the topos
$\SetH{}$. We anticipate that, if this can be done, then
$k(\Ga\SR)\simeq\Ga\kSR$.

As was discussed in Section \eq{SubSec:SR}, the presheaf $\SR$ is
defined at each stage $V$\ by
\begin{equation}
 \SR_V:=\{\l:\downarrow\!\!V\map\mathR
\mid\l\in\mathcal{OR}(\downarrow\!\!V,\mathR)\}.
\end{equation}
If $i_{V^\prime V}:V^\prime\subseteq V$, then the presheaf map
from $\SR_V$ to $\SR_{V^\prime}$ is just the restriction of the
order-reversing functions from $\downarrow\!\! V$ to
$\downarrow\!\! V^\prime$.

The first step in constructing $\kSR$ is to define an equivalence
relation on pairs of functions, $\l,\ka\in\SR_V$, for each stage
$V$, by saying that $(\l_1,\ka_1) \equiv (\l_2,\ka_2)$ if, and
only, there exists $\alpha\in\SR_V$ such that
\begin{equation}
        \l_1(V^\prime)+\ka_2(V^\prime)+\alpha(V^\prime)=
        \ka_1(V^\prime)+\l_2(V^\prime)+\alpha(V^\prime)
\end{equation}
for all $V^\prime\subseteq V$.

{\definition The presheaf  $\kSR$ is defined over the category
$\V{}$ in the following way.
\begin{enumerate}
\item[(i)] On objects $V\in\Ob{\V{}}$:
\begin{equation}
 \kSR_V:=\{[\l,\ka]\mid\l,\ka\in\mathcal{OR}(\downarrow\!\!V,\mathR)\},
\end{equation}
where $[\l,\ka]$ denotes the $k$-equivalence class of $(\l,\ka)$.

\item[(ii)] On morphisms $i_{V^\prime V}:V^\prime\subseteq V$: The
arrow $\kSR(i_{V^\prime V}):\kSR_V\map\kSR_{V^\prime}$ is given by
$\big(\kSR(i_{V^\prime
V})\big)([\l,\ka]):=[\l|_{V^\prime},\ka|_{V^\prime}]$ for all
$[\l,\ka]\in\kSR_V$.
\end{enumerate}
}

It is straightforward to show that $\kSR$ is an abelian
group-object in the topos $\SetH{}$. In particular, an arrow
$+:\kSR\times\kSR\map\kSR$ is defined at each stage $V$ by
\begin{equation}
  +_V\big([\l_1,\ka_1],[\l_2,\ka_2]\big):=[\l_1+\l_2,\ka_1+\ka_2]
\end{equation}
for all
$\big([\l_1,\ka_1],[\l_2,\ka_2]\big)\in\kSR_V\times\kSR_V$. It is
easy to see that (i) $\Ga\kSR\simeq k(\Ga\SR)$; and (ii) $\SR$ is
a sub-object of $\kSR$ in the topos $\SetH{}$.

\subsubsection{The Presheaf $\kSR$ as the Quantity-Value Object.}
We can now identify $\kSR$ as a possible quantity-value object in
$\SetH{}$. To each bounded, self-adjoint operator $\A$, there is
an arrow $[\dasBo{A}]:\Sig\map\kSR$, given by first sending
$\A\in\BH_\sa$ to $\dasBo{A}$ and then taking $k$-equivalence
classes. More precisely, one takes the monic
$\iota:\SR\hookrightarrow\kSR$ and then constructs
$\iota\circ\dasBo{A}:\Sig\map\kSR.$

Since, for each stage $V$, the elements in the image of
$[\dasBo{A}]_V= (\iota\circ\dasBo{A})_V$ are of the form $[\l,0]$,
$\l\in\SR_V$,  their square is well-defined. From a physical
perspective, the use of $\kSR$ rather than $\SR$ renders possible
the definition of things like the `intrinsic dispersion',
$\nabla(\A):=\dasBo{A^2}-\dasBo{A}^2$; see \eq{Def:nabla}.

\paragraph{The relation between $\PR{\mathR}$ and $\kSR$.}
In Section \ref{SubSec:SR}, we considered the presheaf
$\PR{\mathR}$ of order-preserving and order-reversing functions as
a possible quantity-value object. The advantage  of this presheaf
is the symmetric utilisation of inner and outer daseinisation, and
the associated physical interpretation of arrows from $\Sig$ to
$\PR{\mathR}$.

It transpires that $\PR{\mathR}$ is closely related to $\kSR$.
Namely, for each $V$, we define an equivalence relation $\equiv$
on $\PR{\mathR}_V$ by
\begin{equation}\label{EquivRelInPR}
                (\mu_1,\nu_1) \equiv (\mu_2,\nu_2)
\makebox{ iff } \mu_1+\nu_1=\mu_2+\nu_2.
\end{equation}
Then $\PR{\mathR}/\equiv$ is isomorphic to $\kSR$ under the
mapping
\begin{equation}
    [\mu,\nu]\mapsto[\nu,-\mu]\in\kSR_V
\end{equation}
for all $V$ and all $[\mu,\nu]\in(\PR{\mathR}/\equiv)_V$. (Recall
that $[\mu,\nu]\in(\PR{\mathR}/\equiv)_V$ means that $\mu$ is
order-preserving and $\nu$ is order-reversing.)

However, there is a difference between the arrows that represent
physical quantities. The arrow $[\dasBo{A}]:\Sig\map\kSR$ is given
by first sending $\A\in\BH_\sa$ to $\dasBo{A}$ and then taking
$k$-equivalence classes---a construction that only involves outer
daseinisation. On the other hand, there is  an arrow
$[\dasB{A}]:\Sig\map\PR{\mathR}/\equiv$, given by first sending
$\A$ to $\dasB{A}$ and then taking the equivalence classes defined
in (\ref{EquivRelInPR}). This involves both inner and outer
daseinisation.

We can show that $[\dasBo{A}]$ uniquely determines $\A$ (see
Appendix), but currently it is an open question if $[\dasB{A}]$
also fixes $\A$ uniquely.

\section{The Role of Unitary Operators}
\label{Sec:Unitary}
\subsection{The Daseinisation of Unitary Operators}
Unitary operators play an important  role in the formulation of
quantum theory, and we need to understand the analogue of this in
our topos formalism.

Unitary operators arise in the context of both `covariance' and
`invariance'. In elementary quantum theory, the `covariance'
aspect comes the fact that if we have made the associations
\begin{eqnarray*}
\mbox{Physical state }&\mapsto&
        \mbox{state vector $\ket\psi\in\Hi$}\\[5pt]
\mbox{Physical observable }A &\mapsto&
        \mbox{self-adjoint operator $\A$ acting on $\Hi$}
\end{eqnarray*}
then the same physical predictions will be obtained if the
following associations are used instead
\begin{eqnarray}
\mbox{Physical state }&\mapsto&
     \mbox{state vector $\U \ket\psi\in\Hi$} \label{UCovar}\\[5pt]
\mbox{Physical observable }A &\mapsto&
    \mbox{self-adjoint operator }\U\A\U^{-1}
                \mbox{ acting on $\Hi$}\nonumber
\end{eqnarray}
for any unitary operator $\hat U$. Thus the mathematical
representatives of physical quantities are defined only up to
arbitrary transformations of the type above. In non-relativistic
quantum theory, this leads to the canonical commutation relations;
the angular-momentum commutator algebra; and the unitary time
displacement operator. Similar considerations in relativistic
quantum theory involve the Poincare group.

The `invariance' aspect of unitary operators arises when the
operator commutes with the Hamiltonian, giving rise to conserved
quantities.

\paragraph{Daseinisation of unitary operators.} As a side remark,
we first consider the question if daseinisation can be applied to
a unitary operator $\U$. The answer is clearly `yes', via the
spectral representation:
\begin{equation}
        \U=\int_{\mathR}e^{i\l} d\hat E^U_\l
\end{equation}
where $\l\mapsto E^\U_\l$ is the spectral family for $\U$. Then,
in analogy with \eqs{Def:dastooVA}{Def:dastoiVA} we have the
following: {\definition The \emph{outer daseinisation},
$\dastoo{}U$, resp.\ the \emph{inner daseinisation}, $\dastoi{}U$,
of a unitary operator $\U$ are defined as follows:
\begin{eqnarray}
\dastoo{V}{U}&:=&\int_\mathR e^{i\l}\,
       d\big(\delta^i_V(\hat E^U_{\l}) \big),\label{Def:dastooVU}
       \\
       \dastoi{V}{U}&:=&\int_{\mathR}e^{i\l}\, d
       \big(\bigwedge_{\mu>\l}\delta^o_V(\hat E^U_{\mu})\big),
\label{Def:dastoiVU}
\end{eqnarray}
at each stage $V$. }

To interpret these entities\footnote{It would be possible to
`complexify' the presheaf $\kSR$ in order to represent unitary
operators as arrows from $\Sig$\ to $\mathC\kSR$.  However, there
is no obvious physical use for this procedure.} we need to
introduce a new presheaf defined as follows.

{\definition The {\em outer, unitary de Groote presheaf}, $\dOU$,
is defined by:
\begin{enumerate}
\item[(i)] On objects $V\in\Ob{\V{}}$:  $\dOU_V:=V_{\rm un}$, the
collection of unitary operators in $V$.

\item[(ii)] On morphisms $i_{V^{\prime}V}:V^{\prime }\subseteq V:$
The mapping $\dOU(i_{V^{\prime}\, V}):\dOU_V
\map\dOU_{V^{\prime}}$ is given by
\begin{eqnarray}
        \dOU(i_{V^{\prime}\, V})(\hat\alpha)&:=&
                \dastoo{V^{\prime}}{\alpha}\\
&=&\int_\mathR e^{i\l}\,  d
\big(\delta^i(\hat E^\alpha_\l)_{V^{\prime}}\big)\\
&=&\int_\mathR e^{i\l}\,  d\big(\H(i_{V^\prime\,V})(\hat
E^\alpha_\l)\big)
\end{eqnarray}
for all $\hat\alpha\in\dOU_V$.
\end{enumerate}
} \noindent Clearly, (i) there is an analogous definition of an
`inner', unitary de Groote presheaf; and (ii) the map $V\mapsto
\dastoo{V}{U}$ defines a global element of $\dOU$.

This definition has the interesting consequence that, at each
stage $V$,
\begin{equation}
        \delta^o(e^{i\A})_V=e^{i\dastoo{V}{A}}
\end{equation}
A particular example of this construction is the one-parameter
family of unitary operators, $t\mapsto e^{it\hat H}$, where $\hat
H$ is the Hamiltonian of the system.

Of course, in our case everything commutes. Thus suppose
$g\mapsto\U_g$ is a representation of a Lie group $G$ on the
Hilbert space $\Hi$. Then these operators can be daseinised to
give the map $g\mapsto\delta^o(\U_g)$, but generally this is not a
representation of $G$ (or of its Lie algebra) since, at each stage
$V$ we have
\begin{equation}
 \delta^o(\U_{g_1})_V\delta^o(\U_{g_2})_V =
 \delta^o(\U_{g_2})_V\delta^o(\U_{g_1})_V
\end{equation}
for all $g_1,g_2\in G$. Clearly, there is an analogous result for
inner daseinisation.

\subsection{Unitary Operators and Arrows in $\SetH{}$.}
\subsubsection{The Definition of $\ell_{\hat U}:\Ob{\V{}}\map\Ob{\V{}}$}
In classical physics, the analogue of unitary operators are
`canonical transformations'; \ie\ symplectic diffeomorphisms from
the state space $\S$ to itself. This suggests that should try to
associate arrows in $\SetH{}$ with each unitary operator $\hat U$.

Thus we want to see if unitary operators can  act on the objects
in $\SetH{}$. In fact, if $\UH$ denotes the group of all unitary
operators in $\Hi$, we would like to find a \emph{realisation} of
$\UH$ in the topos $\SetH{}$.

As a first step, if $\U\in\UH$ and $V\in\Ob{\V{}}$ is an abelian
von Neumann subalgebra of $\BH$, let us define
\begin{equation}
        \ell_\U(V):=\{\U\A\U^{-1}\mid \A\in V\}.
                                \label{Def:ellU(V)}
\end{equation}
It is clear that $\ell_\U(V)$ is a unital, abelian algebra of
operators, and that self-adjoint operators are mapped into
self-adjoint operators.  Furthermore, the map
$\A\mapsto\U\A\U^{-1}$ is continuous in the weak-operator
topology, and hence, if $\{\A_i\}_{i\in I}$ is a weakly-convergent
net of operators in  $V$, then $\{\U\A_i\U^{-1}\}_{i\in I}$ is a
weakly-convergent net of operators in $\ell_\U(V)$, and vice
versa.

It follows that $\ell_\U(V)$ is an abelian von Neumann algebra
(\ie\ it is weakly closed), and hence $\ell_\U$ can be viewed as a
map $\ell_\U:\Ob{\V{}}\map\Ob{\V{}}$.  We note the following:
\begin{enumerate}
\item Clearly, for all $\U_1,\U_2\in\UH$,
\begin{equation}
        \ell_{\U_1}\circ\ell_{\U_2}=\ell_{\U_1\U_2}
        \label{RepUHV(H)}
\end{equation}
Thus $\U\mapsto\ell_\U$ is a realisation of the group $\UH$ as a
group of transformations of $\Ob{\V{}}$.

\item For all $U\in\UH$, $V$ and $\ell_\U(V)$ are
\emph{isomorphic} subalgebras of $\BH$, and
$\ell_\U^{-1}=\ell_{\U^{-1}}$.

\item If $V^\prime\subseteq V$, then, for all $\U\in\UH$,
\begin{equation}
        \ell_\U(V^\prime)\subseteq\ell_\U (V).
\end{equation}
Hence, each transformation $\ell_\U$ preserves the
partial-ordering of the poset category $\V{}$.

From this it follows  that each $\ell_\U:\Ob{\V{}}\map\Ob{\V{}}$
is a \emph{functor} from the category $\V{}$ to itself.

\item One consequence of the order-preserving property of
$\ell_\U$ is as follows. Let  $S$ be a sieve of arrows on $V$,
\ie\ a collection of subalgebras of $V$ with the property that if
$V^\prime\in S$, then, for all $V^{\prime\prime}\subseteq
V^\prime$ we have $V^{\prime\prime}\in S$. Then
\begin{equation}
        \ell_\U(S):=\{\ell_\U(V^\prime)\mid V^\prime\in S\}
        \label{Def:ellU(S)}
\end{equation}
is a sieve of arrows on $\ell_\U(V)$.\footnote{In the partially
ordered set $\V{}$, an arrow from $V^{\prime}$ to $V$ can be
identified with the subalgebra $V^{\prime}\subseteq V$, since
there is exactly one arrow from $V^{\prime}$ to $V$.}
\end{enumerate}

\subsubsection{The Effect of $\ell_\U$ on Daseinisation}
We recall that if $\P$ is any projection, then the (outer)
daseinisation, $\dastoo{V}{P}$, of $\P$ at stage $V$ is
\begin{equation}
\dastoo{V}{P}:=\bigwedge\big\{\hat{Q}\in\mathcal{P}(V)\mid
\hat{Q}\succeq \P\big\},\label{Def:dasouter2}
\end{equation}
where we have resorted once more to using the propositional
language $\PL{S}$. Thus
\begin{eqnarray}
        \U\dastoo{V}{P}\U^{-1}&=&\U\bigwedge
\big\{\hat{Q}\in\mathcal{P}(V)\mid\hat{Q}\succeq \P\big\}\U^{-1} \nonumber\\
        &=& \bigwedge\big\{\U\hat{Q}\U^{-1}\in\mathcal{P}
(\ell_\U(V))\mid\hat{Q}\succeq \P\big\} \nonumber\\
        &=& \bigwedge\big\{\U\hat{Q}\U^{-1}\in\mathcal{P}
(\ell_\U(V))\mid\U\hat{Q}\U^{-1}\succeq \U\P\U^{-1}\big\} \nonumber\\
        &=& \daso(\U\P\U^{-1})_{\ell_\U(V)}             \label{Udas(P)U_1=}
\end{eqnarray}
where we used the fact that the map $\hat Q\mapsto\U\hat Q\U^{-1}$
is weakly continuous.

Thus we have the important result
\begin{equation}
\U\dastoo{V}{P}\U^{-1}=\daso(\U\P\U^{-1})_{\ell_\U(V)}
                \label{Utransdas}
\end{equation}
for all unitary operators $\U$, and for all stages $V$. There is
an analogous result for inner daseinisation.

Equation \eq{Utransdas} can be applied to the de Groote presheaf
$\dG$ to give
\begin{equation}
\U\dastoo{V}{A}\U^{-1}=\delta^{o}(\U\A\U^{-1})_{\ell_\U(V)}
\end{equation}
for unitary operators $\U$, and all stages $V$.

We recall that the truth sub-object, $\TO^{\ket\psi}$, of the
outer presheaf, $\G$, is defined at each stage $V$ by
\begin{eqnarray}
 \TO^{\ket\psi}_V&:=&\{\hat\alpha\in \G_V\mid
                {\rm Prob}(\hat\alpha;\ket\psi)=1\}\label{TOpsi1}
                                                \nonumber\\[2pt]
        &=&\{\hat\alpha\in \G_V\mid
                \bra\psi\hat\alpha\ket\psi=1\}     \label{TOpsi2}
\end{eqnarray}
The neo-realist, physical interpretation of $\TO^{\ket\psi}$ is
that the `truth' of the proposition represented by $\hat P$  is
\begin{eqnarray}
\TVal{\dasoB{P}\in\TO}{\ket\psi}_V &:=&\{V^\prime\subseteq V\mid
\dastoo{V^\prime}{P}\in\TO^{\ket\psi}_{V^\prime}\}  \\
        &=&\{V^\prime\subseteq V\mid \bra\psi\dastoo{V^\prime}{P}
                        \ket\psi=1\}\label{TVDas}
\end{eqnarray}
for all stages $V$. We then get
\begin{eqnarray}
\hspace{-1.4cm}
 \ell_\U\big(\TVal{\dasoB{P}\in\TO}{\ket\psi}_V\big)
\hspace{-0.5cm} &\overset{\eq{TVDas}}{=}&
\ell_\U\{V^\prime\subseteq V\mid\bra\psi
        \dastoo{V^\prime}{P}\ket\psi=1\}\\
&=& \{\ell_\U(V^\prime)\subseteq\ell_\U(V)\mid\bra\psi
                        \dastoo{V^\prime}{P}\ket\psi=1\}\\
&=& \{\ell_\U(V^\prime)\subseteq\ell_\U(V)\mid\bra\psi
                \U^{-1}\U\dastoo{V^\prime}{P}\U^{-1}\U\ket\psi=1\}\\
&\overset{\eq{Utransdas}}{=}& \{\ell_\U(V^\prime)\subseteq
\ell_\U(V)\mid\bra\psi\U^{-1}\daso(\U\P\U^{-1})_{\ell_\U(V)}
                                            \U\ket\psi=1\}\\
       &=& \TVal{\daso(\U\P\U^{-1})\in\TO}{\U\ket\psi}_{\ell_\U (V).}
\end{eqnarray}
Thus we get the important result
\begin{equation}
\TVal{\daso(\U\P\U^{-1})\in\TO} {\U\ket\psi}_{\ell_\U(V)}=
        \ell_\U\big(\TVal{\dasoB{P}\in\TO}{\ket\psi}_V\big).
                \label{[]UCovar}
\end{equation}
This can be viewed as the topos analogue of the statement in
\eq{UCovar} about the invariance of the results of quantum theory
under the transformations $\ket\psi\mapsto\U\ket\psi$,
$\A\mapsto\U\A\U^{-1}$.

\subsubsection{The $\U$-twisted Presheaf}
Let us return once more to the definition \eq{Def:ellU(V)} of the
functor $\ell_\U:\V{}\map\V{}$. As we shall see in the next paper,
\cite{DI(4)}, any such functor induces a `geometric morphism' from
$\SetH{}$ to $\SetH{}$. The exact definition is not needed here:
it suffices to remark that part of this geometric morphism is an
arrow $\ell_\U^*:\SetH{}\map\SetH{}$ defined by
\begin{equation}
        \ps{F}\mapsto\ell_\U^*\ps{F}:=\ps{F}\circ\ell_\U.
\end{equation}

Note that, if $\U_1,\U_2\in\UH$ then, for all presheaves $\ps{F}$,
\begin{eqnarray}
        \ell_{\U_2}^*(\ell_{\U_1}^*\ps{F})&=&
                (\ell_{\U_1}^*\ps{F})\circ\ell_{\U_2}=
        (\ps{F}\circ\ell_{\U_1})\circ\ell_{\U_2}\nonumber\\
        &=&\ps{F}\circ(\ell_{\U_1}\circ\ell_{\U_2})=
                        \ps{F}\circ\ell_{\U_1\U_2}\nonumber\\
&=&\ell_{\U_1\U_2}^*\ps{F}.
\end{eqnarray}
Since this is true for all functors $\ps{F}$ in $\SetH{}$, we
deduce that
\begin{equation}
        \ell_{\U_2}^*\circ\ell_{\U_1}^*=\ell_{\U_1\U_2}^*
\end{equation}
and hence the map $\U\mapsto\ell_\U^*$ is an (anti-)representation
of the group $\UH$ by arrows in the topos $\SetH{}$.

Of particular interest to us are the presheaves $\ell_\U^*\Sig$
and $\ell_U ^*\kSR$. We denote them by $\Sig^\U$ and $\kSR^\U$
respectively and say that they are `$\U$-twisted'.

{\theorem For each $\U\in\UH$, there is a natural isomorphism
$\iota:\Sig\map\Sig^\U$ as given in the following diagram
\begin{center}
\setsqparms[1`1`1`1;1000`700]
\square[\Sig_V`\Sig_V^\U`\Sig_{V^{\prime}}`\Sig_{V^\prime}^\U;
\iota^\U_V`\Sig(i_{V^\prime V})`\Sig^\U(i_{V^\prime
V})`\iota^\U_{V^\prime}]
\end{center}
where, at each stage $V$,
\begin{equation}
        (\iota^U_V(\l))(\A):=\l(\U^{-1}\A\U)
\end{equation}
for all $\l\in\Sig_V$, and all $\A\in V_{\sa}$.}

The proof, which just involves chasing round the diagram above
using the basic definitions, is not included here.

Even simpler is the following theorem: {\theorem For each
$\U\in\UH$, there is a natural isomorphism
$\ka^\U:\SR\map(\SR)^\U$ whose components
$\ka^\U_V:\SR_V\map(\SR)^\U_V$ are given by
\begin{equation}
\ka^\U_V(\mu)(\ell_\U(V^\prime)):=\mu(V^\prime) \label{Def:kRRU}
\end{equation}
for all $V^\prime\subseteq V$.}

Here, we recall $\mu\in\SR_V$ is a function
$\mu:\downarrow\!\!V\map\mathR$ such that if $V_2\subseteq
V_1\subseteq V$ then $\mu(V_2)\geq\mu(V_1)$, \ie\ an
order-reversing function. In \eq{Def:kRRU} we have used the fact
that there is a bijection between the sets
$\downarrow\!\!\ell_\U(V)$ and $\downarrow\!\!V$.

Finally, {\theorem We have the following commutative diagram:
\begin{center}
                \setsqparms[1`1`1`1;1000`700]
                \square[\Sig`\Sig^\U`\SR`\SR^\U.;
                \iota^\U`\dasB{A}`\breve{\delta}(\U^{-1}\A\U)`\ka^\U]
\end{center}
}

\paragraph{The analogue of unitary operators for a general topos.}
It is  interesting to reflect on  the analogue of the above
constructions for a general topos. It soon becomes  clear that,
once again, we encounter the antithetical concepts of `internal'
and `external'.

For example, in the discussion above, the unitary operators and
the group $\UH$ lie outside the topos $\SetH{}$ and enter directly
from the underlying, standard quantum formalism. As such, they are
external to both the languages $\PL{S}$ and $\L{S}$.  We
anticipate that notions of `covariance' and `symmetry'  have
applications well beyond those in classical physics and quantum
physics. However,  at the very least, in a general topos one would
presumably replace the external $\UH$ with an internal group
object in the topos concerned. And, of course,  the notion of
`symmetry' is closely related to the concept to time, and time
development, which opens up a Pandora's box of possible
speculation. These issues are important, and await further
development.

\section{Conclusion}
In this, the third in our series of papers on topos theory and
physics, we have completed the work begun in the second paper
where we showed how propositions can be represented by clopen
sub-objects of the spectral presheaf, $\Sig$, of  toposified
quantum theory. This is equivalent to finding a topos
representation of the propositional language $\PL{S}$, and is the
natural completion of the earlier work on quantum theory and
topoi.

In the present paper we have gone well beyond this earlier work by
finding a representation of the full local language $\L{S}$. In
particular, we identified the quantity-value object as being the
presheaf $\SR$; or, possibly, its $k$-extension $\kSR$ or,
$\PR{\mathR}$. This  enabled us to find a representation of a
physical quantity, $A$, with an arrow $\dasBo{A}:\Sig\map\SR$.

However, we reiterate  that our main interest is \emph{not} to
find a new interpretation of quantum theory, but   rather to
develop a general framework in which new types of theories of
physics can be developed; and in which both classical and quantum
physics  arise as special cases.

It follows  that an important challenge for future work is to show
that our general topos scheme can be used to develop genuinely new
theories of physics, not just to rewrite old ones in a new
language. Of particular  interest  is the problem with which we
motivated the scheme in the first place: namely, to find tools for
constructing theories that go beyond quantum theory and  which do
not use Hilbert spaces, path integrals, or any of the other
familiar entities in which the continuum real and/or complex
numbers play a fundamental role.

As we have discussed, the topoi for quantum systems are of the
form $\SetH{}$, and hence  embody contextual logic in a
fundamental way. One way of going `beyond' quantum theory, while
escaping the \emph{a priori} imposition of continuum concepts, is
to use presheaves over a more general `category of contexts',
$\mathcal C$, \ie \ develop the theory in the topos
$\Set^{{\mathcal C}^{{\rm op}}}$. Such a structure embodies
contextual, multi-valued logic in an intrinsic way, and in that
sense might be said to encapsulate one of the fundamental insights
of quantum theory. However, and unlike in quantum theory, there is
no obligation to use the real or complex numbers in the
construction of the category $\mathcal C$.

Of course, although true, this remark does not of itself give a
concrete example of a theory of this type. However, it is
certainly a pointer in a novel direction, and one at which we
would not have arrived  if the challenge to `go beyond quantum
theory' had been construed only in terms of trying to generalise
Hilbert spaces, path integrals, and the like.

From a more general perspective, other types of topoi are possible
realms for the construction of physical theories. One
 simple, but mathematically rich example arises from
the theory of $M$-sets. Here, $M$ is a monoid and, like all
monoids, can be viewed as a category with a single object, and
whose arrows are  the elements of $M$.  Thought of as a category,
a  monoid is `complementary' to  a partially-ordered set. In a
monoid, there is only one object, but plenty of arrows from that
object to itself; whereas in a partially-ordered set there are
plenty of objects, but at most one arrow between any pair of
objects.  Thus a partially-ordered set is  the most economical
category with which to capture the concept of `contextual logic'.
On the other hand, the logic associated with a monoid is
non-contextual as there is only one object in the category.

It is easy to see that a functor from $M$ to $\Set$ is just an
`$M$-set': \ie\  a set on which $M$  acts as a monoid of
transformations. An arrow between two such $M$-sets is  an
equivariant map between them. In physicists' language, one would
say that the topos $\Set^M$---usually denoted $BM$--- is the
category of the `non-linear realisations' of $M$.

The sub-object classifier, $\O_{BM}$,  in  $BM$ is the collection
of left ideals in $M$; hence, many of the important constructions
in the topos can be handled using the algebraic language.  The
topos $BM$ is one of the simplest to define and work with and, for
that reason, it is a popular source of examples in texts on topos
theory. It would be intriguing to experiment with constructing
model theories of physics using one of these simple topoi. One
possible use of $M$-sets is discussed in \cite{Isham05b} in the
context of reduction of the state vector, but there will surely be
others.

\noindent {\bf Acknowledgements} This research was supported by
grant RFP1-06-04 from The Foundational Questions Institute
(fqxi.org).  AD gratefully acknowledges financial support from the
DAAD.

This work is also supported in part by the EC Marie Curie Research
and Training Network ``ENRAGE'' (European Network on Random
Geometry) MRTN-CT-2004-005616.

\section*{Appendix}

We will show that the mapping
\begin{eqnarray}
\theta:\BH_\sa &\map& \rm{Hom}_{\SetH{}}(\Sig,\SR)\\
             \A &\mapsto& \dasBo{A}
\end{eqnarray}
from self-adjoint operators in $\mathcal{B(H)}$ to the natural
transformations/arrows from the spectral presheaf $\Sig$ to the
quantity-value object $\SR$, is injective.

Let $\nu:\Sig\map\SR$ be an arrow in the image of $\theta$. Then
there is some self-adjoint operator $\A$ such that
$\nu=\dasBo{A}$. Our task is to show that $\A$ is unique. Let
$\P\in\PH$, and let $V_{\P}:=\{\P,\hat{1}\}^{\prime\prime}$ be the
abelian von Neumann algebra generated by $\P$. Let
$\widetilde{H}_{\P}:=\{\P,\hat {1}\}$ be the ultrafilter in
$\mathcal{P}(V_{\P})$ corresponding to the character
$\l_{\P}\in\Sig_{V_{\P}}$ (\ie\
$\l_{\P}(\P)=\l_{\P}(\hat{1})=\hat{1}$). The cone over
$\widetilde{H}_{\P}$ in $\mathcal{P(H)}$ is
\begin{equation}
\mathcal{C}(\widetilde{H}_{\P})=H_{\P}=\{\hat{Q}
\in\mathcal{P(H)}\mid\hat{Q}\succeq\P\},
\end{equation}
which is the principal filter in $\PH$ generated by
$\P\in\mathcal{P(H)}$. We have
\begin{eqnarray}
                \dasBVo{V_{\P}}{A}(\l_{\P})(V_{\P})
                &=& \l_{\P}(\dastoo{V_{\P}}{A})\\
                &=& f_{\dastoo{V_{\P}}{A}}(\widetilde{H}_{\P})\\
                &=& f_{\A}(\mathcal{C}(\widetilde{H}_{\P}))\\
                &=& f_{\A}(H_{\P}).
\end{eqnarray}
Here, we used (\ref{Eq_f_doVA(F_lambda)=lambda(doVA)}) and
(\ref{Eq_f_dastooVA(I)=f_A(C(I))}). Letting $\P$ vary over $\PH$,
we get the values of the observable function $f_{\A}$ at all
principal filters $H_{\P}$.

De Groote has shown \cite{deG05b} that there is a bijection
\begin{eqnarray}
                \omega:\BH_\sa &\map& \mathcal{O(D(H))}\\
                \A &\mapsto& f_{\A}
\end{eqnarray}
between self-adjoint operators and observable functions
$f_{\A}:\mathcal{D(H)}\map\mathR$. Moreover, each observable
function $f_{\A}$ is uniquely determined by its restriction to
principal filters. This shows that, given an arrow $\nu$ in the
image of $\theta$, we can reconstruct from it a \emph{unique}
self-adjoint operator $\A\in\BH_\sa$.

Now let
\begin{equation}
     [\dasBo{A}]:\Sig\map\kSR
\end{equation}
denote the natural transformation from the spectral presheaf to
the abelian group-object $\kSR$, given by first sending $\A$ to
$\dasBo{A}$ and then taking the $k$-equivalence classes at each
stage $V$. The monoid $\SR$ is embedded into $\kSR$ by sending
$\mu\in\SR_V$ to $[\mu,0]\in\kSR_V$ for all $V$, which implies
that $\A$ is also uniquely determined by
$[\dasBo{A}]$.\footnote{In an analogous manner, one can show that
the arrows $\dasBi{A}:\Sig\map\OP$ and $[\dasBi{A}]:\Sig\map
k(\OP)$ uniquely determine $\A$, and that the arrow
$\dasB{A}:\Sig\map\PR{\mathR}$ also uniquely determines $\A$. It
is currently unknown if the arrow
$[\dasB{A}]:\Sig\map\PR{\mathR}/\equiv$ also fixes $\A$ uniquely.}

On the level of the local language $\L{S}$, we will assume that
$\L{S}$ contains function symbols $A:\Sigma\map\mathcal R$
corresponding bijectively to a set of self-adjoint operators
$\A\in\BH_\sa$. Hence, these function symbols are faithfully
represented by the natural transformations $\dasBo{A}$.

Interestingly, these results all carry over to  an arbitrary
unital von Neumann algebra $\mathcal N\subseteq\BH$. In this way,
the formalism is flexible enough to adapt to situations where we
have symmetries (which can described mathematically by a von
Neumann algebra $\mathcal N$ that has a non-trivial commutant) and
super-selection rules (which corresponds to $\mathcal N$ having a
non-trivial centre).

\end{document}